\documentstyle[12pt,psfig,hpatex]{article}  

\newcommand{\ra}{\rightarrow}

\newcommand{\be}{\begin{equation}}
\newcommand{\ee}{\end{equation}}
\newcommand{\bea}{\begin{eqnarray}}
\newcommand{\eea}{\end{eqnarray}}
\newtheorem{thm}{Theorem}[section]

\def\bbbr{{\rm I\!R}} 

\def\bbbn{{\rm I\!N}} 

\def\bbbm{{\rm I\!M}}

\def\bbbh{{\rm I\!H}}

\def\bbbf{{\rm I\!F}}

\def\bbbk{{\rm I\!K}}

\def\bbbl{{\rm I\!L}}

\def\bbbp{{\rm I\!P}}

\def\bbbe{{\rm I\!E}}

\def\bbbone{{\mathchoice {\rm 1\mskip-4mu l} {\rm 1\mskip-4mu l}
{\rm 1\mskip-4.5mu l} {\rm 1\mskip-5mu l}}}

\def\bbbc{{\mathchoice {\setbox0=\hbox{$\displaystyle\rm C$}\hbox{\hbox
to0pt{\kern0.4\wd0\vrule height0.9\ht0\hss}\box0}}
{\setbox0=\hbox{$\textstyle\rm C$}\hbox{\hbox
to0pt{\kern0.4\wd0\vrule height0.9\ht0\hss}\box0}}
{\setbox0=\hbox{$\scriptstyle\rm C$}\hbox{\hbox
to0pt{\kern0.4\wd0\vrule height0.9\ht0\hss}\box0}}
{\setbox0=\hbox{$\scriptscriptstyle\rm C$}\hbox{\hbox
to0pt{\kern0.4\wd0\vrule height0.9\ht0\hss}\box0}}}}

\def\bbbq{{\mathchoice {\setbox0=\hbox{$\displaystyle\rm Q$}\hbox{\raise
0.15\ht0\hbox to0pt{\kern0.4\wd0\vrule height0.8\ht0\hss}\box0}}
{\setbox0=\hbox{$\textstyle\rm Q$}\hbox{\raise
0.15\ht0\hbox to0pt{\kern0.4\wd0\vrule height0.8\ht0\hss}\box0}}
{\setbox0=\hbox{$\scriptstyle\rm Q$}\hbox{\raise
0.15\ht0\hbox to0pt{\kern0.4\wd0\vrule height0.7\ht0\hss}\box0}}
{\setbox0=\hbox{$\scriptscriptstyle\rm Q$}\hbox{\raise
0.15\ht0\hbox to0pt{\kern0.4\wd0\vrule height0.7\ht0\hss}\box0}}}}
\def\bbbt{{\mathchoice {\setbox0=\hbox{$\displaystyle\rm
T$}\hbox{\hbox to0pt{\kern0.3\wd0\vrule height0.9\ht0\hss}\box0}}
{\setbox0=\hbox{$\textstyle\rm T$}\hbox{\hbox
to0pt{\kern0.3\wd0\vrule height0.9\ht0\hss}\box0}}
{\setbox0=\hbox{$\scriptstyle\rm T$}\hbox{\hbox
to0pt{\kern0.3\wd0\vrule height0.9\ht0\hss}\box0}}
{\setbox0=\hbox{$\scriptscriptstyle\rm T$}\hbox{\hbox
to0pt{\kern0.3\wd0\vrule height0.9\ht0\hss}\box0}}}}

\def\bbbs{{\mathchoice
{\setbox0=\hbox{$\displaystyle     \rm S$}\hbox{\raise0.5\ht0\hbox
to0pt{\kern0.35\wd0\vrule height0.45\ht0\hss}\hbox
to0pt{\kern0.55\wd0\vrule height0.5\ht0\hss}\box0}}
{\setbox0=\hbox{$\textstyle        \rm S$}\hbox{\raise0.5\ht0\hbox
to0pt{\kern0.35\wd0\vrule height0.45\ht0\hss}\hbox
to0pt{\kern0.55\wd0\vrule height0.5\ht0\hss}\box0}}
{\setbox0=\hbox{$\scriptstyle      \rm S$}\hbox{\raise0.5\ht0\hbox
to0pt{\kern0.35\wd0\vrule height0.45\ht0\hss}\raise0.05\ht0\hbox
to0pt{\kern0.5\wd0\vrule height0.45\ht0\hss}\box0}}
{\setbox0=\hbox{$\scriptscriptstyle\rm S$}\hbox{\raise0.5\ht0\hbox
to0pt{\kern0.4\wd0\vrule height0.45\ht0\hss}\raise0.05\ht0\hbox
to0pt{\kern0.55\wd0\vrule height0.45\ht0\hss}\box0}}}}

\def\bbbz{{\mathchoice {\hbox{$\sf\textstyle Z\kern-0.4em Z$}}
{\hbox{$\sf\textstyle Z\kern-0.4em Z$}}
{\hbox{$\sf\scriptstyle Z\kern-0.3em Z$}}
{\hbox{$\sf\scriptscriptstyle Z\kern-0.2em Z$}}}}

\def\bC{\ifmmode{\bbbc }\else${\bbbc }$\fi}
\def\bE{\ifmmode{\bbbe }\else${\bbbe }$\fi}
\def\bF{\ifmmode{\bbbf }\else${\bbbf }$\fi}
\def\bH{\ifmmode{\bbbh }\else${\bbbh }$\fi}
\def\bI{\ifmmode{\bbbone }\else${\bbbone }$\fi}
\def\bK{\ifmmode{\bbbk }\else${\bbbk }$\fi}
\def\bL{\ifmmode{\bbbl }\else${\bbbl }$\fi}
\def\bM{\ifmmode{\bbbm }\else${\bbbm }$\fi}
\def\bN{\ifmmode{\bbbn }\else${\bbbn }$\fi}
\def\bP{\ifmmode{\bbbp }\else${\bbbp }$\fi}
\def\bQ{\ifmmode{\bbbq }\else${\bbbq }$\fi}
\def\bR{\ifmmode{\bbbr }\else${\bbbr }$\fi}
\def\bS{\ifmmode{\bbbs }\else${\bbbs }$\fi}
\def\bZ{\ifmmode{\bbbz }\else${\bbbz }$\fi}

\def\cA{\ifmmode{\cal A}\else${\cal A}$\fi}
\def\cB{\ifmmode{\cal B}\else${\cal B}$\fi}
\def\cC{\ifmmode{\cal C}\else${\cal C}$\fi}
\def\cD{\ifmmode{\cal D}\else${\cal D}$\fi}
\def\cE{\ifmmode{\cal E}\else${\cal E}$\fi}
\def\cF{\ifmmode{\cal F}\else${\cal F}$\fi}
\def\cG{\ifmmode{\cal G}\else${\cal G}$\fi}
\def\cH{\ifmmode{\cal H}\else${\cal H}$\fi}
\def\cI{\ifmmode{\cal I}\else${\cal I}$\fi}
\def\cJ{\ifmmode{\cal J}\else${\cal J}$\fi}
\def\cK{\ifmmode{\cal K}\else${\cal K}$\fi}
\def\cL{\ifmmode{\cal L}\else${\cal L}$\fi}
\def\cM{\ifmmode{\cal M}\else${\cal M}$\fi}
\def\cN{\ifmmode{\cal N}\else${\cal N}$\fi}
\def\cO{\ifmmode{\cal O}\else${\cal O}$\fi}
\def\cP{\ifmmode{\cal P}\else${\cal P}$\fi}
\def\cQ{\ifmmode{\cal Q}\else${\cal Q}$\fi}
\def\cR{\ifmmode{\cal R}\else${\cal R}$\fi}
\def\cS{\ifmmode{\cal S}\else${\cal S}$\fi}
\def\cT{\ifmmode{\cal T}\else${\cal T}$\fi}
\def\cU{\ifmmode{\cal U}\else${\cal U}$\fi}
\def\cV{\ifmmode{\cal V}\else${\cal V}$\fi}
\def\cW{\ifmmode{\cal W}\else${\cal W}$\fi}
\def\cX{\ifmmode{\cal X}\else${\cal X}$\fi}
\def\cY{\ifmmode{\cal Y}\else${\cal Y}$\fi}
\def\cZ{\ifmmode{\cal Z}\else${\cal Z}$\fi}

\def\bk#1{{\ifmmode{\langle#1\rangle}\else${\langle#1\rangle}$\fi}} 

\def\sp#1#2{{\ifmmode{\langle{#1}|{#2}\rangle}
\else${\langle{#1}|{#2}\rangle}$\fi}}

\def\prf{\medbreak\noindent{\bf Proof.}\enspace}

\def\qed{\hfill $\sqcap$\llap{$\sqcup$}\break}

\def\data{\the\day\space\ifcase\month\or January \or February \or March \or
April \or May \or June \or July \or August \or September
\or October \or November \or December \fi\space\the\year}

\def\date{\the\day\space\ifcase\month\or Janvier \or F\'evrier \or Mars \or
Avril \or Mai \or Juin \or Juillet \or Ao\^ut \or Septembre
\or Octobre \or Novembre \or D\'ecembre \fi\space\the\year}

\def\sc{\scriptstyle}

\def\t{{\tt t}}
\def\W{{\tt W}}

\def\writefig#1 #2 #3 {\rlap{\kern #1 truecm \raise #2 truecm
\hbox{#3}}} \def\figtext#1{\smash{\hbox{#1}} \vspace{-5mm}}
\def\legende#1{\caption{\protect \small #1}}

\begin{document}
\title{MATHEMATICAL THEORY OF THE WETTING\\[2mm]
       PHENOMENON IN THE 2D ISING MODEL{$\,^{i}$}}
\authors{Charles-Edouard Pfister}
\address{D\'epartement de Math\'ematiques, EPF-L \\
CH-1015 Lausanne, Switzerland\\
e-mail: cpfister@eldp.epfl.ch}
\authors{and Yvan Velenik{$\,^{ii}$}}
\address{D\'epartement de Physique, EPF-L\\
CH-1015 Lausanne, Switzerland\\
e-mail:velenik@eldp.epfl.ch}
\abstract{ 
We give a
mathematical theory of the wetting phenomenon in the 2D Ising
model using the formalism of Gibbs states. We treat the grand
canonical and  canonical ensembles.}

\renewcommand{\thefootnote}{\roman{footnote}}
\footnotetext[1]{Published in Helv.Phys.Acta {\bf 69}, 949--973 (1996).
Dedicated to K.Hepp and W.Hunziker.}
\renewcommand{\thefootnote}{\arabic{footnote}}
\footnotetext[2]{Supported by Fonds National Suisse
Grant 2000-041806.94/1}

\section{Introduction}\label{Introduction}
\setcounter{equation}{0}

We study the wetting phenomenon in the 2D Ising model, starting
from basic principles of Statistical Mechanics.  The results of
section \ref{grandens.} are based on \cite{FP1}, \cite{FP2} and
\cite{FP3} and  those of section \ref{canens.} follow from
recent results on the large deviations of the magnetization
\cite{PV}. We shall in general refer to these papers for proofs.
Our purpose is to give a global view of the mathematical
results, which are now fairly complete. Since the results about
large deviations  are valid in the 2D case only we restrict the
whole discussion to this case.

Let us suppose that we have a binary mixture and that the
physical parameters are chosen so that we have coexistence of
the two phases, called $+$ phase and $-$ phase. The system is
inside a box; the horizontal bottom wall $w$ of the box adsorbs
preferentially the $-$ phase. If we prepare the system in the
$+$ phase we may observe the formation of a thin film of the $-$
phase between the wall $w$ and the $+$ phase in the bulk, so
that the $+$ phase cannot be in contact with this wall. This is
the phenomenon of complete wetting of the wall. Notice that the
total amount of the $-$ phase is not fixed a priori. The
appropriate statistical ensemble to describe this situation is a
grand canonical ensemble. The surface tension
$\widehat{\tau}(\theta)$ is the surface free energy due to an
interface between the two coexisting phases, making an angle
$\theta$ with an horizontal reference line. The contribution to
the surface free energy due to the wall $w$ when the bulk phase
is the $+$ phase is $\tau^+$. In the case of complete wetting of
$w$ we expect that $\tau^{+}$ can be decomposed into
$\tau^{-}+\widehat{\tau}(0)$, where  $\tau^{-}$ is the surface
free energy due to the wall $w$ in presence of the $-$ phase and
$\widehat{\tau}(0)$ is the surface tension of an horizontal
interface between the film of the $-$ phase and the bulk $+$
phase. On the other hand, if the wall adsorbs preferentially the
$+$ phase, then we expect that \be
\tau^{+}=\tau^{-}-\widehat{\tau}(0)\,. \ee Indeed, if we impose
the $-$ phase in the bulk, then we create an interface between
the film of the $+$ phase near the wall and the $-$ bulk phase.
When complete wetting does not occur a stability argument shows
that one expects  a strict inequality \be |\tau^{+}-\tau^{-}|<
\widehat{\tau}(0)\,. \ee In such a case  the bulk phase is in
contact with the wall. In other words the state of the system
near the wall depends on the nature of the bulk phase.
Therefore, depending on the property of the wall, we have
\be\label{cahncrit} |\tau^{+}-\tau^{-}|\leq \widehat{\tau}(0)\,,
\ee with equality if and only if complete wetting holds. This
fundamental relation has been derived in a thermodynamical
setting by Cahn \cite{C}. In \cite{FP1}, \cite{FP2} and
\cite{FP3} this situation is analysed in the Ising model and the
fundamental inequality (\ref{cahncrit}) is derived directly from
the microscopic hamiltonian within the standard setting of
Statistical Mechanics. The criterion for complete wetting, \be
|\tau^{+}-\tau^{-}|= \widehat{\tau}(0)\,, \ee is interpreted in
terms of unicity of the (surface) Gibbs state.

It is possible to imagine another situation, which is often used
to present the phenomenon of wetting of a wall. We have again
the same system prepared in the $+$ phase, but we put a
macroscopic droplet of the $-$ phase inside the box. The droplet
is attached to the wall $w$ if the wall adsorbs preferentially
the $-$ phase. The shape of the droplet depends on the
interactions between the wall and the binary mixture; the shape
of the droplet is given by the solution of an isoperimetric
problem with constraint \cite{Wi} (Winterbottom's construction).
Since the total amount of the $-$ phase is macroscopic and
fixed, the relevant ensemble is  a canonical ensemble. The case
of complete wetting of the wall $w$ corresponds to the total
spreading of the droplet of $-$ phase against the wall. The
relation between the shape of the droplet and criterion
(\ref{cahncrit}) is made through the contact angle $\theta$ between
the droplet and the wall (see Fig. \ref{fig_angles}), \be\label{herring}
\cos(\theta)\cdot
\widehat{\tau}(\theta)-\sin(\theta)\cdot
{d\widehat{\tau}(\theta)\over d\theta} =\tau^{+}-\tau^{-}\,, \ee
the so--called Herring-Young equation. A derivation of
(\ref{herring}) is given in section \ref{isoperimetric}. Let us
consider the following two extreme (degenerate) situations. When
the wall adsorbs preferentially the $+$ phase, then \be
\tau^{+}-\tau^{-}= -\widehat{\tau}(0)<0\,. \ee This corresponds
in (\ref{herring}) to $\theta=\pi$, which means that the droplet
of $-$ phase is not attached to the wall. On the other hand,
when the wall adsorbs preferentially the $-$ phase, then \be
\tau^{+}-\tau^{-}= \widehat{\tau}(0)>0\,, \ee which corresponds
in (\ref{herring}) to the case $\theta=0$. This means that the
droplet does not exist as such, but spreads out completely
against the wall. We show that the above situations can be
rigorously derived for the 2D Ising model in an appropriate
canonical ensemble. The analysis consists in deriving  sharp
estimates of the large deviations of the magnetization
\cite{PV}. Our work \cite{PV}  extends considerably previous
works \cite{MS1}, \cite{MS2}, \cite{DKS}, \cite{Pf2}, \cite{I1}
and \cite{I2} on the subject, since we can treat the case of an
arbitrary boundary magnetic field.

\section{Ising model}\label{ising} \setcounter{equation}{0}

\subsection{Gibbs states}\label{gibbs}

The lattice is \be \bZ^2:=\{\,t=(t(1),t(2)):\,t(i)\in\bZ\,\}.
\ee A {\bf spin configuration} is a function $\omega$ defined on
$\bZ^2$, $t\mapsto\omega(t)$, with $\omega(t)=\pm 1$. The Ising
variable at $t$ is \be \sigma(t)(\omega):=\omega(t)\,. \ee An
{\bf edge} of $\bZ^2$, $\bk{t,t'}$, is  a pair of nearest
neighbours sites of the lattice $\bZ^2$. We also call edge the
unit--length segment in $\bR^2$ with end--points $t,t'$. For
each edge $\bk{t,t'}$ we have a coupling constant
$J(\bk{t,t'})\geq 0$. For each finite subset
$\Lambda\subset\bZ^2$ the energy is \be
H_\Lambda:=-\sum_{\bk{t,t'}\cap\Lambda\not=\emptyset}
J(\bk{t,t'})\sigma(t)\sigma(t')\,. \ee Let $\omega^*$ be given;
the Gibbs measure in $\Lambda$ with boundary condition
$\omega^*$ and inverse temperature $\beta$ is the probability
measure \be \mu_{\Lambda}^{\omega^*,\beta}(\omega):=\cases{
Z^{\omega^*}(\Lambda)^{-1}\exp(-\beta H_{\Lambda}(\omega))& if
$\omega(t)=\omega^*(t)$ for all $t\not\in\Lambda$\cr
0&otherwise.\cr} \ee The normalization constant is called {\bf
partition function}.

Assume that all coupling constants are equal to one. If we
choose $\omega^*$ such that $\omega^*(t)=1$ for all $t$, then
there is a limiting measure \be \mu^{+,\beta}:=
\lim_{\Lambda\uparrow\bZ^2}\mu^{\omega^*,\beta}_{\Lambda}\,. \ee
The same is true for $\omega^*$, such that $\omega^*(t)=-1$ for
all $t$. The limiting measure is $\mu^{-,\beta}$. These two
Gibbs states are translation invariant and extremal. There is a
unique Gibbs measure if and only if
$\mu^{+,\beta}=\mu^{-,\beta}$. This happens if and only if
$\beta\leq\beta_c$, where $\beta_c$ is the inverse critical
temperature. The inverse critical temperature is characterized
by the property that there is a positive spontaneous
magnetization, \be
m^*(\beta):=\int\sigma_t(\omega)\mu^{+,\beta}(d\omega)>0 \ee if
and only if $\beta>\beta_c$.

\subsection{Contours}

\begin{figure}[htb]
\centerline{\psfig{figure=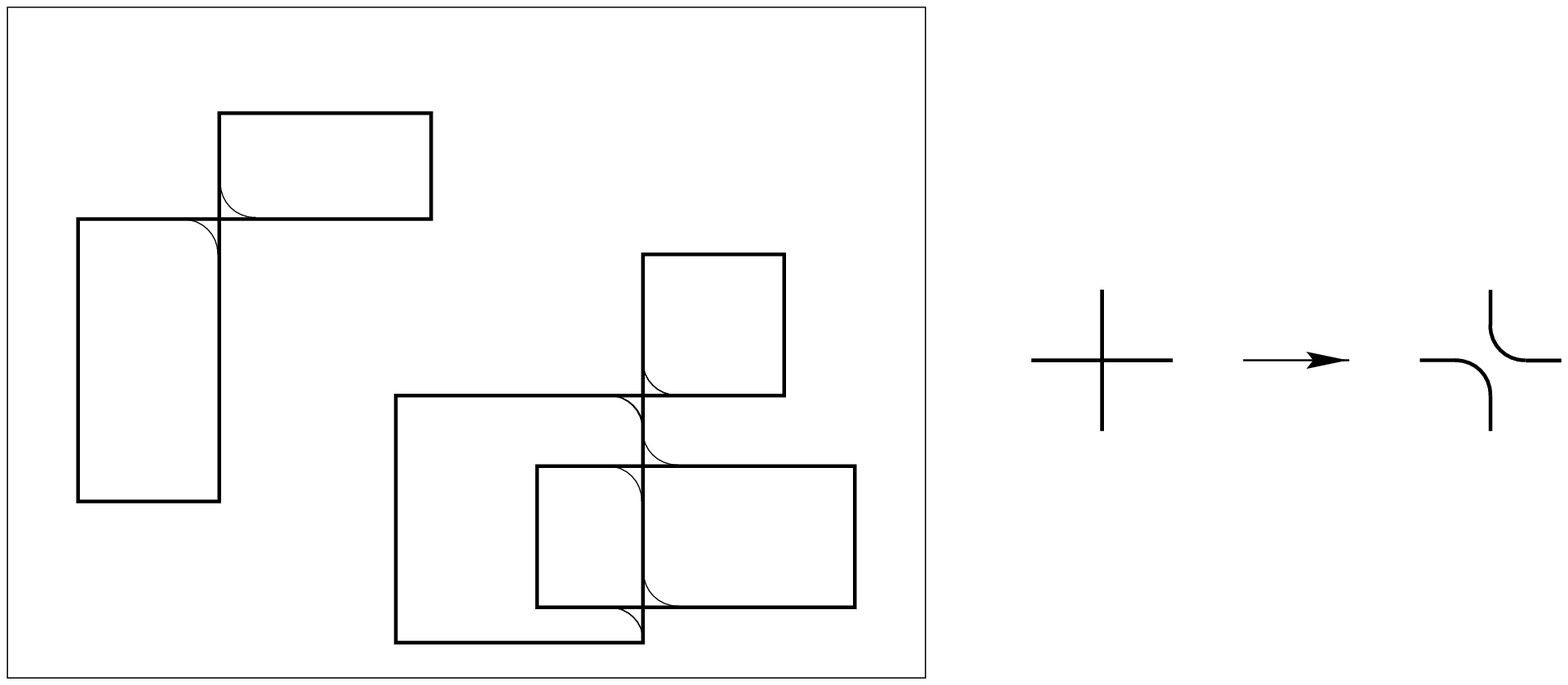,height=50mm}} \figtext{
\writefig	10.3	1.5	{Deformation rule} } \legende{Two connected
components giving rise to four contours} \label{fig_contours}
\end{figure}

Let $\omega^*$ be a fixed boundary condition. The usual way of
describing the configurations of the model is to specify the
pairs of nearest neighbours sites $e=\bk{t,t'}$ such that
$\sigma(t)\sigma(t')=-1$. Equivalently we specify the dual edges
$e^*$, that is the edges of the {\bf dual lattice} $(\bZ^2)^*$,
\be (\bZ^2)^*:=\{\,t=(t(1),t(2)):\,t(i)+1/2\in\bZ\,\} \ee which
cross the edges $e$. We decompose the set formed by all these
dual edges into connected components. In \cite{PV} we further
decompose the connected components into a set of paths, called
${\bf contours}$, using the rule given in Figure \ref{fig_contours}; for
details see \cite{PV}. As set of edges the paths are disjoint
two by two. Some paths are closed and are called {\bf closed
contours}; some are open and are called {\bf open contours}.

\section{Grand canonical ensemble, Cahn's
criterion}\label{grandens.} \setcounter{equation}{0}

\subsection{Surface Gibbs states}\label{surface}

Since we are interested in  boundary effects, we consider the
Ising model on the following rectangular box
$\Lambda_L(r_1,r_2)=\Lambda_L$. Let $r_1, r_2\in\bN$; we set
\be\label{box} \Lambda_L(r_1,r_2):=\{t\in\bZ^2:\,-r_1L\leq t(1)<
r_1L\,;\; 0\leq t(2)<2r_2 L\}\,. \ee The coupling constants of
the model are \be\label{coupling} J(\bk{t,t'}):=\cases{ h & if
$t=(s,-1)$ and $t'=(s,0)$ with $s\in\bZ$,\cr 1 & otherwise.\cr}
\ee We consider four different boundary conditions \be
\omega^*(t):=\cases{ a & $a=\pm 1$, if $t(2)<0$,\cr b & $b=\pm
1$, if $t(2)\geq 0$.\cr} \ee The hamiltonian $H_L\equiv
H_{\Lambda_L}$ with $(a,b)$ boundary condition can be written
\be H_{L}=-\sum_{{\sc \bk{t,t'}:\atop\sc t,t'\in\Lambda_L}}
\sigma(t)\sigma(t') -\sum_{{\sc t\in\Lambda_L\,:\atop\sc
t(2)=0}}ha\,\sigma(t) -\sum_{{\sc t\in\Lambda_L\,:t(2)=2r_2
L\atop\sc {\rm or}\; t(1)=\pm r_1L }}b\,\sigma(t)\,. \ee The
term \be -\sum_{{\sc t\in\Lambda_L\,:\atop\sc
t(2)=0}}ha\,\sigma(t) \ee describes the interaction of the
binary mixture inside $\Lambda_L$ with the bottom wall of the
box $\Lambda_L$, which plays the role of the wall $w$. The wall
adsorbs preferentially the $a$ phase. We interpret $ha$ as a
real--valued boundary magnetic field and we refer to the
$b$--part of the boundary condition, \be\label{2.6} -\sum_{{\sc
t\in\Lambda_L\,:t(2)=2r_2 L\atop\sc {\rm or}\; t(1)=\pm r_1L
}}b\,\sigma(t)\,, \ee as the boundary condition. Expectation
value with respect to the Gibbs measure in $\Lambda_L$ is
denoted by $\bk{\,\cdot\,}_L^{b}(\beta,h)$. When $h\geq 0$ and
$b=1$, then all contours of a configuration are closed. On the
other hand, when $h<0$ and $b=1$, then there is exactly one open
contour\footnote{Although there is a close connection between
the open contour $\lambda$ and the interface, which is created
by the bulk $+$ phase and the wall which adsorbs preferentially
the $-$ phase, one must not identify the open contour with the
interface.} in each configuration, denoted below by $\lambda$,
with end--points $t_{l,L}=(-r_1L-1/2,-1/2)$ and
$t_{r,L}=(r_1L-1/2,-1/2)$. The boundary condition (\ref{2.6})
specifies the type $b$ of the phase in the bulk of the system in
the following sense \footnote{ For positive $h$ this is proven
in \cite{FP2}; the result is already valid if we replace the
condition $t(2)\geq L^{1/2+\delta}$ by $t(2)\geq d$. The same
result holds for $h>-h^*$ defined in subsection \ref{cahn} using
Lemma 7.1 of \cite{PV}. We need condition $t(2)\geq
L^{1/2+\delta}$ only in the case of complete wetting of the wall
by the $-$ phase. The proof is based on the following simple
fact. The surface tension $\widehat{\tau}(\theta)$ is smooth and
has a minimum at $\theta=0$; If $\theta=L^{-1/2+\delta}$, then
$L\widehat{\tau}(\theta)\sim L\widehat{\tau}(0)+
L/2\widehat{\tau}''(0)L^{-1+2\delta}= L\widehat{\tau}(0)+
1/2\widehat{\tau}''(0)L^{2\delta}$. Lemma 5.5 in \cite{PV}
implies that the probability that the open contour $\lambda$
visits sites $t$, $t(2)\geq L^{1/2+\delta}$, goes to zero faster
than $L^2\exp\{-O(L^{2\delta})\}$. }. Let $\beta>\beta_c$ and
$\delta>0$. Given $\varepsilon>0$ there exist $L_0$ and $d$ such
that, $\forall L\geq L_0$,
\be
|\bk{\sigma(t)}_L^{+}(\beta,h)-m^*(\beta)\,|\leq\varepsilon\,,
\ee 
for all $t=(t(1),t(2))\in\Lambda_L$ satisfying the
conditions 
\be |t(1)-r_1L|\geq d\quad,\quad |t(2)-2r_2L|\geq d
\quad,\quad t(2)\geq L^{1/2+\delta}\,. 
\ee

The {\bf surface Gibbs states} are the limiting measures \be
\bk{\,\cdot\,}^b(\beta,h):=
\lim_{L\ra\infty}\bk{\,\cdot\,}_L^b(\beta,h)\,, \ee or limiting
measures defined by choosing  different boundary conditions. The
existence of the limits when $b=\pm $, as well as the following
properties are proven in \cite{FP1}. \begin{enumerate} \item The
two states $\bk{\,\cdot\,}^b(\beta,h)$, $b=\pm $, are extremal
Gibbs states. \item They are invariant under the translations
$x\mapsto x+y$, $y=(y(1),0)$. \item There is a unique surface
Gibbs measure if and only if
$\bk{\,\cdot\,}^+(\beta,h)=\bk{\,\cdot\,}^-(\beta,h)$.
\end{enumerate}

\subsection{Surface tension and surface free energies }

We define the basic thermodynamic quantities which enter in the
description of the wetting phenomenon. More details on the
surface tension are given in \cite{Pf1} and \cite{Pf2} section
6. Surface free energies are studied in \cite{FP1} and
\cite{FP2}; see also \cite{PV}. The definitions we use are
standard. Explanations for them are given in \cite{Pf1}. These
definitions are not very satisfactory from a conceptual point of
view (see beginning of the introduction of \cite{ABCP}), since
they are defined without using a precise notion of interface.
This weak point is also a strong point, because the notion of
interface is a delicate notion, which is difficult to analyse.

\subsubsection{Surface tension}

\begin{figure}[t]
\centerline{\psfig{figure=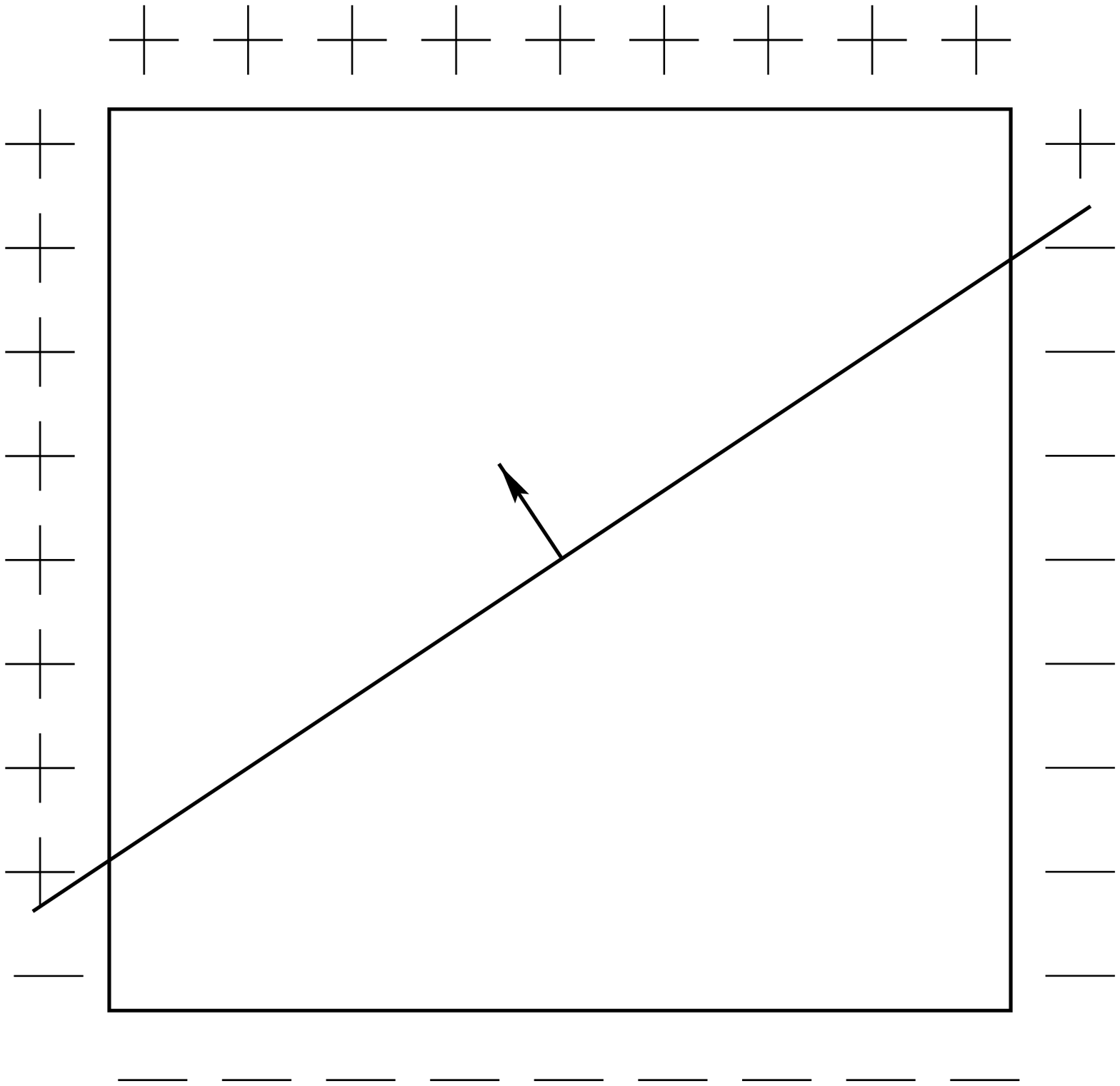,height=50mm}}
\figtext{ 
\writefig	9.6	1.1	{$\Omega_L$}
\writefig	7.7	2.9	{$n$} 
\writefig	6.2	1.1	{$\cL(n)$} }
\legende{Definition of the $\omega^n$ boundary condition}
\label{fig_surfacetension} 
\end{figure}

For latter purposes (see section \ref{isoperimetric}) it is more
convenient to parametrize surface tension  using the normal
vector to the interface instead of the angle $\theta$. We
consider the model with coupling constants equal to one on the
whole lattice. Let $\Omega_L$ be the square box \be
\Omega_L:=\{t\in\bZ^2:\,-L\leq t(1),t(2)\leq L\}\,. \ee Let
$n=(n(1),n(2))$ be a unit vector in $\bR^2$ and $\cL(n)$ a line
perpendicular to $n$, passing through $(0,0)$. We define a boundary
condition $\omega^n$ by setting (see Fig. \ref{fig_surfacetension})
\be 
\omega^n(t):=\cases{1& if
$t$ is above or on the line $\cL(n)$,\cr -1& if $t$ is below the
line $\cL(n)$.\cr} 
\ee 
The corresponding partition function is
$Z^{\omega^n}(\Omega_L)$. By definition, {\bf the surface
tension of an interface perpendicular to $n$} is \be
\widehat{\tau}(n):=\lim_{L\ra\infty}-{1\over 2L}
\ln{Z^{\omega^n}(\Omega_L)\over Z^{+}(\Omega_L)}\,. \ee 

We usually do not write explicitly the $\beta$--dependence of
the surface tension. We extend the function $\widehat{\tau}$ to
$\bR^2$ as a positively homogeneous function. By
GKS--inequalities it follows that the function is subadditive,
\be
\widehat{\tau}(x+y)\leq\widehat{\tau}(x)+\widehat{\tau}(y)\,.
\ee It can be shown that $\widehat{\tau}$ is strictly positive
(for all $x\not=0$) if and only if $\beta>\beta_c$ \cite{LP}.
(This is also true for dimensions higher than two.)

{\bf Remark:} In dimension two the surface tension is related by
duality to the decay--rate of the two--point function at the
dual inverse temperature \cite{BLP1}, \cite{Pf2}. The
decay--rate of the two--point function at the dual inverse
temperature can be computed explicitly  \cite{MW}.  The surface
tension has the following symmetry properties, \be
\widehat{\tau}(n)=\widehat{\tau}(-n)\quad{\rm and}\quad
\widehat{\tau}(n)=\widehat{\tau}(m)\quad{\rm if}\quad n\perp
m\,. \ee It is a smooth function for any $\beta<\infty$.

\subsubsection{Surface free energies}\label{freeenergies}

The coupling constants are given by (\ref{coupling}). Free
energies are defined up to an arbitrary constant. We need only
to define $\tau^- - \tau^+$, the difference of the contributions
of the wall $w$ to the free energy, when the bulk phase is the
$-$ phase respectively the $+$ phase. The definition is similar
to the definition of the surface tension. Let $\Lambda_L$ be the
box (\ref{box}). By definition, \be \tau^- - \tau^+:=\lim_{L\ra
\infty}-{1\over 2r_1L}\ln{Z^-(\Lambda_L)\over Z^+(\Lambda_L)}\,.
\ee We usually do not write explicitly the
$(\beta,h)$--dependence of $\tau^- - \tau^+$. The existence of
the limit is proven in \cite{FP2}, as well as the following
results. \begin{enumerate} \item For any $\beta$, $|\tau^- -
\tau^+|\leq\widehat{\tau}((1,0))$. Hence $\tau^- - \tau^+=0$ for
all $\beta\leq\beta_c$. \item We have the symmetry \be \tau^-(h)
- \tau^+(h)=-\Big(\tau^-(-h) - \tau^+(-h)\Big)\,. \ee \item Let
$\beta>\beta_c$ and $n_w:=(0,-1)$. For positive $h$, $\tau^-(h)
- \tau^+(h)$ is a positive concave function; for all $h\geq 1$
\be \tau^-(h) - \tau^+(h)=\widehat{\tau}(n_w)\,. \ee
\end{enumerate} These results are not restricted to dimension
two.

{\bf Remark:} In dimension two the quantity $\tau^- - \tau^+$
can be computed \cite{P}. For $h$ positive it is equal to the
decay--rate of the boundary two--point function at the dual
coupling constants.

\subsection{Cahn's criterion and phase diagram}\label{cahn}

In all the section the inverse critical temperature
$\beta>\beta_c$ is fixed. We study the model in the grand
canonical ensemble; the coupling constants are given by
(\ref{coupling}) and the states are the surface Gibbs states
$\bk{\,\cdot\,}^{b}$ of subsection \ref{surface}. For
definiteness we choose the $+$  boundary condition, so that we
have the $+$  phase as bulk phase.

\begin{figure}[t]
\centerline{\psfig{figure=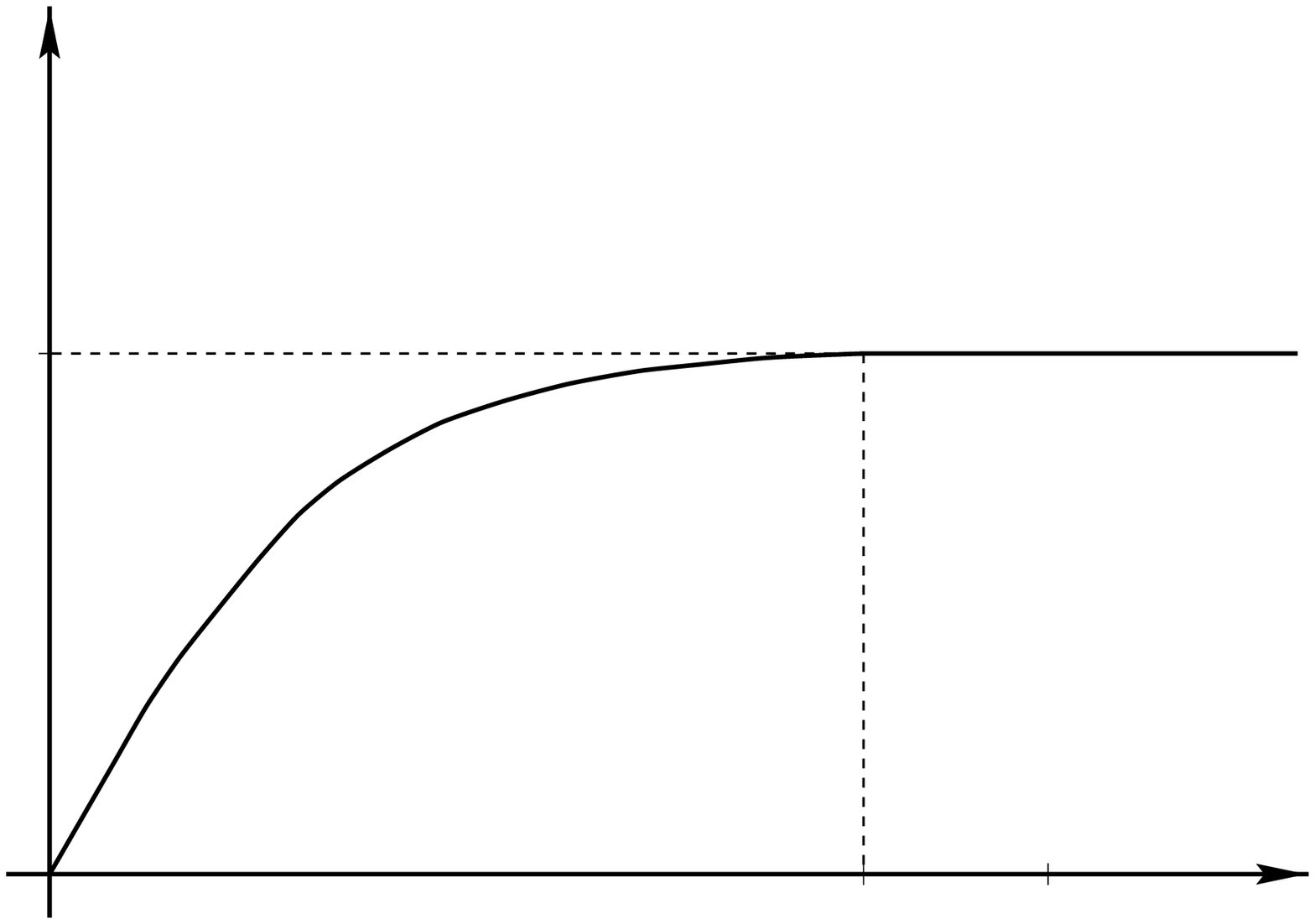,height=50mm}} \figtext{
\writefig	3.21	4.70	{$\tau^--\tau^+$} 
\writefig      11.15	0.30	{$h$} 
\writefig	9.15	0.30	{$h^*$}
\writefig	10.14	0.30	{$1$}
\writefig	3.50	3.50	{$\widehat\tau(n_w)$} }
\legende{$\tau^--\tau^+$ as a function of $h$.}
\label{fig_taupm} \end{figure}

From the properties of $\tau^--\tau^+$,  which are stated in
subsection \ref{freeenergies}, we have the existence of a
positive $h^*(\beta)$ such that (see Fig. \ref{fig_taupm})
\be 
h^*(\beta):=\inf\Big\{h\geq
0:\, \tau^-(\beta,h)-\tau^+(\beta,h)=
\widehat{\tau}(n_w)(\beta)\Big\}\,. 
\ee
The thermodynamical criterion of Cahn states that there is complete wetting of
the wall if and only if $|h|\geq h^*(\beta)$.

{\bf Remark:} The value of $h^*(\beta)$ is known in dimension
two. It was computed by Abraham \cite{A}; $h^*(\beta)$ is the
solution of the equation \be \exp\{2\beta\} \{\cosh 2\beta-\cosh
2\beta h^*(\beta)\}= \sinh 2\beta\,. \ee In \cite{FP3}, see also
\cite{PP}, it is shown that the value of $h^*(\beta)$ can also
be obtained from the work of McCoy and Wu \cite{MW}. 
 
\begin{figure}
\centerline{\psfig{figure=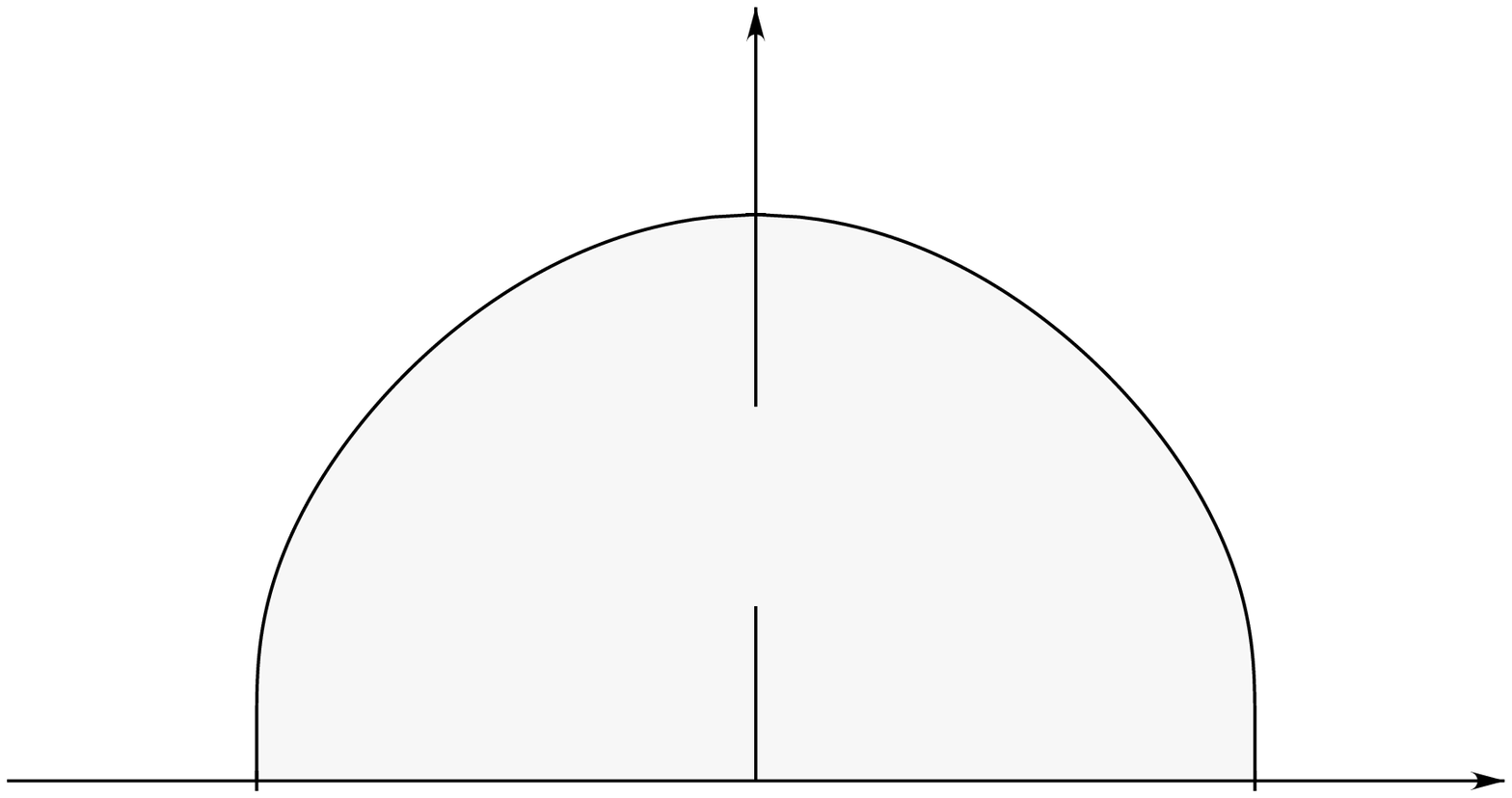,height=50mm}}
\figtext{ 
\writefig	7.7	4.9	{$T$}
\writefig	8.35	4.25	{$T_c$} 
\writefig	12.5	0.1	{$h$}
\writefig	11.2	0.1	{$1$} 
\writefig	4.8	0.1	{$-1$}
\writefig	6.5	2.5	{Non-uniqueness of}
\writefig	6.5	2.0	{surface Gibbs state} } \legende{Phase
diagram. The region of non-uniqueness of the surface Gibbs state
is shaded. In the other region, there is a single surface Gibbs
state.} \label{fig_phdiag} \end{figure}

\bigskip To make the
connection with the surface Gibbs states we use the identity
proven in \cite{FP2} \be \tau^-(\beta,h)
-\tau^+(\beta,h)=\int_0^h\Big(\bk{\sigma(0)}^+(\beta,h')-
\bk{\sigma(0)}^-(\beta,h')\Big)\,dh'\,. \ee Therefore \bea
\widehat{\tau}(n_w)&=& \tau^-(\beta,h^*) -\tau^+(\beta,h^*)\\
&=&\int_0^{h^*} \Big(\bk{\sigma(0)}^+(\beta,h')-
\bk{\sigma(0)}^-(\beta,h')\Big)\,dh'\,.\nonumber \eea For all
$h\geq h^*$ we must have \be \bk{\sigma(0)}^+(\beta,h)=
\bk{\sigma(0)}^-(\beta,h)\,, \ee since for positive $h'$ \be
\bk{\sigma(0)}^+(\beta,h')\geq \bk{\sigma(0)}^-(\beta,h')\,. \ee
On the other hand, if $0\leq h< h^*$, the concavity of
$\tau^-(\beta,h)-\tau^+(\beta,h)$ as function of $h$ implies
that \be \bk{\sigma(0)}^+(\beta,h)> \bk{\sigma(0)}^-(\beta,h)\,.
\ee These results, together with the properties of the surface
Gibbs states mentionned in subsection \ref{surface}, imply the
following interpretation of the Cahn's criterion for complete
wetting (see also the corresponding phase diagram of Fig. \ref{fig_phdiag}).

\begin{enumerate} 
\item There is complete wetting of the wall $w$ if and only if there is a
unique surface Gibbs state. 
\item
There is a unique surface Gibbs state if and only if 
\be
|\tau^-(\beta,h)-\tau^+(\beta,h)|=\widehat{\tau}(n_w)(\beta)>0\,.
\ee 
\item The value $h^*$ of the boundary magnetic field where the wetting
transition takes place can be defined as 
\be
h^*(\beta):=\inf_{h\geq 0}\Big\{\bk{\,\cdot\,}^+(\beta,h)=
\bk{\,\cdot\,}^-(\beta,h)\Big\}\,. 
\ee 
\end{enumerate} 
{\bf Remarks:} 1.
These results are proven in \cite{FP2}; they are not restricted
to dimension two.

2. In dimension two we have precise information about the
behaviour of the open contour $\lambda$ entering in the
description of the configurations when we have $-$ boundary
condition and $0\leq h< h^*$. The contours sticks to the wall.
There exists a constant $K$, such that the probability that the
open contour $\lambda$ does not visit the segment
$I=\{t:\,t(2)=0, x_1\leq t(1) \leq x_2\}$ is smaller than \be
K\exp\{-\kappa\cdot|x_1-x_2|\}\,, \ee with \be
\kappa=\widehat{\tau}(n_w)- \Big(\tau^--\tau^+\Big)\,. \ee This
result is a combination of Lemma 7.1 in \cite{PV} and the remark
following the proof of that lemma. By symmetry the same result
applies to the case where we have $+$ boundary condition and
$0\geq h>-h^*$.

3. Weaker results about the behaviour of $\lambda$, but not
restricted to dimension two, are contained in \cite{FP2}.

\section{Variational problem}\label{isoperimetric}
\setcounter{equation}{0}

We consider here the variational problem giving the shape of the
macroscopic droplet when the surface tension and the surface
free energies of the wall are known. In section \ref{canens.} we
show how this variational problem arises when the analysis
starts from the hamiltonian of the model.

The variational problem, which gives the shape of the
macroscopic droplet in presence of a wall, is a generalization
of the classical isoperimetric problem. In the physics
literature the solution of the problem is known as
Winterbottom's construction \cite{Wi}. Wulff's construction
\cite{Wu} corresponds to the special case when the wall has no
effect on the droplet. Dinghas \cite{D} gave a geometrical proof
of Wulff's construction,  which has been extended by Taylor (see
her review \cite{T} for original references). Wulff's solution
and Bonnesen's inequalities, which describe a (strong) stability
property of the solution, are discussed in details in
\cite{DKS}. A completely different proof, valid in the 2D case
only, is given in \cite{DP}\footnote{These few references are
far from complete. Good reviews about the isoperimetric problem
and related topics are \cite{O1} and \cite{O2}.}.  The many
facets of statistical mechanics of equilibrium shapes are
reviewed in \cite{RW} and \cite{Z}.

\subsection{Geometry of the boundary of a convex body}

Let $C\subset\bR^2$ be a compact convex body, that is, a compact
convex set with a non--empty interior. We denote by $\partial C$
the boundary of $C$ and by $(x|y)$ the Euclidean scalar product
of $x,y\in\bR^2$. The {\bf support function } of $C$,
$\tau_{C}$, is defined on $\bR^2$ by \be\label{support}
\tau_{C}(y):=\sup_{x\in C} (x|y)\,. \ee It is immediate from the
definition (\ref{support}) that $\tau_{C}$ is positively
homogeneous and subadditive, \be \tau_{C}(y_1+y_2)\leq
\tau_{C}(y_1)+\tau_{C}(y_2)\,. \ee Since $C$ is compact it is
also clear that for any $y$ there exists $x_y\in C$, in fact
$x_y\in\partial C$, with \be \tau_{C}(y)=(x_y|y)\,. \ee Let
$n\in\bR^2$ be of norm $\|n\|=1$. Define $A(n)$ to be the
hyperplane \be A(n):=\{x\in\bR^2:\,(x|n)=\tau_{C}(n)\}\,. \ee
Then $A(n)$ is a {\bf support plane} for $C$ at $x_n$, that is
$x_n\in A(n)$ and \be C\subset
\{x\in\bR^2:\,(x|n)\leq\tau_{C}(n)\}\,. \ee Conversely, if
$x\in\partial C$, then there exists a support plane for $C$ at
$x$. This is a consequence of the separation result: if $O$ is
an open convex set and $L$ an affine subset, such that $O\cap
L=\emptyset$, then there exists a hyperplane $H$ with the
properties \cite{E} \be L\subset H\quad{\rm and}\quad O\cap
H=\emptyset\,. \ee For $n\in\bR^2$, $\|n\|=1$, $\tau_{C}(n)$
gives the (signed) distance of the hyperplane $A(n)$ to the
origin. The distance is positive if and only if
$0\in\{x\in\bR^2:\,(x|n)\leq\tau_{C}(n)\}$.

\begin{thm}\label{geometry1}

Let $C$ be a compact convex body in $\bR^2$. Let $\tau_{C}$ be
its support function. Then \bea
C&=&\{x\in\bR^2:\,(x|n)\leq\tau_{C}(n)\; \forall n\,,
\|n\|=1\}\\ &=&
\bigcap_{n:\,\|n\|=1}\{x\in\bR^2:\,(x|n)\leq\tau_{C}(n)\}\,.\nonumber
\eea \end{thm}

\prf It is clear that \be C\subset
\bigcap_{n:\,\|n\|=1}\{x\in\bR^2:\,(x|n)\leq\tau_{C}(n)\}\,. \ee
Suppose that $y\not\in C$. We can separate strictly a closed
convex set $B$ and a compact convex set $K$ by a hyperplane,
when they are disjoint \cite{E}. Therefore there exists a
hyperplane \be H=\{x\in\bR^2:\,(x|m)=\delta\}\,, \ee $\|m\|=1$,
such that for all $x\in C$ we have $(x|m)<\delta$ and at the
same time $(y|m)>\delta$. Therefore \be \sup_{x\in
C}(x|m)=\tau_{C}(m)\leq\delta\,, \ee and consequently \be
y\not\in\bigcap_{n:\,\|n\|=1}\{x\in\bR^2:\,(x|n)\leq\tau_{C}(n)\}\,.
\ee \qed

\begin{figure}[t]
\centerline{\psfig{figure=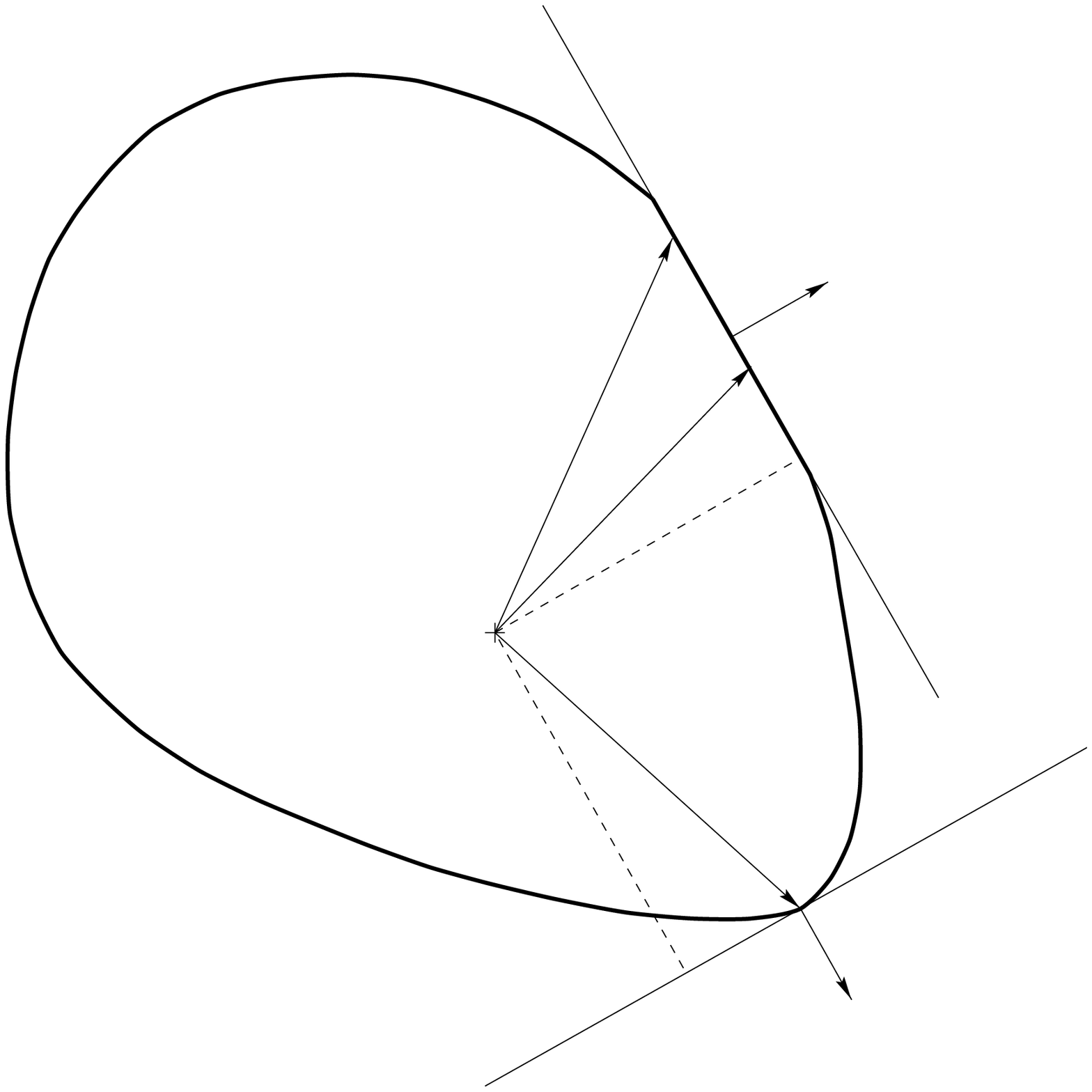,height=70mm}}
\figtext{ 
\writefig	4.4	3.1	{$C$} 
\writefig	7.3	3.1	{$O$}
\writefig	8.85	5.2	{$x_1$} 
\writefig	9.47	1.95	{$x_2$}
\writefig	8.5	4.8	{$y_1$} 
\writefig	7.9	5.0	{$y_2$}
\writefig	8.25	3.05	{grad$\tau(n_2)$} 
\writefig	9.2	5.7	{$n_1$}
\writefig	10.1	1.3	{$n_2$} 
\writefig	9.8	4.5	{$P_1$}
\writefig	8.8	0.8	{$P_2$}
\writefig	8.3	7.1	{$A(n_1)$} 
\writefig	6.8	0.75	{$A(n_2)$} }
\legende{Convex body $C$ with support planes $A(n_1)$, $A(n_2)$.
$x_2$ is a regular point and $A(n_2)$ is a regular support
plane. $x_1$ is a regular point and $A(n_1)$ is not regular. The
vectors $y_1$ and $y_2$ are in the subdifferential of $\tau$ at
$n_1$. dist$(O,P_1)=\tau(n_1)$; dist$(O,P_2)=\tau(n_2)$.}
\label{fig_convexbody} \end{figure}

We recall two definitions (see Fig. \ref{fig_convexbody}). A point $y\in\partial C$ is a {\bf
regular point} if there is a single support plane containing
$y$. This is equivalent to say that the intersection of all
support planes containing $y$ is a $1$--dimensional affine set.
At a regular point $x$ there is a well--defined (unit) normal
vector to $C$, defined as the vector $n$ such that $x\in  A(n)$.
A support plane $A$ is a {\bf regular support plane} if $A\cap
C$ is $0$--dimensional. 

\begin{thm}\label{geometry2} Let $C$ be a compact convex body.
Let $n\in\bR^2$, $\|n\|=1$. Then $A(n)\cap C$ coincides with the
subdifferential $\partial\tau_{C}(n)$ of $\tau_{C}$ at $n$, that
is, \be A(n)\cap
C=\{x\in\bR^2:\,\tau_{C}(z+n)\geq\tau_{C}(n)+(z|x)\quad \forall
z\in\bR^2\}\,. \ee \end{thm}

\prf Suppose that $x\in\partial\tau_{C}(n)$, \be
\tau_{C}(z+n)\geq\tau_{C}(n)+(z|x)\;\forall z\in\bR^2\,. \ee
Since $\tau_{C}$ is subadditive, \be
\tau_{C}(z)+\tau_{C}(n)\geq\tau_{C}(n)+(z|x)\,, \ee we have \be
\tau_{C}(z)\geq (z|x)\quad\forall z\,; \ee hence $x\in C$ by
Theorem \ref{geometry1}. On the other hand, \be
(z+n|x)-\tau_{C}(z+n)\leq (n|x)-\tau_{C}(n)\,; \ee therefore \be
\tau^*_{C}(x)=\sup_{y\in\bR^2}\Big\{\,(y|x)-\tau_{C}(y)\,\Big\}=
(n|x)-\tau_{C}(n)\,. \ee Since $x\in C$, we have for all $y$ \be
(y|x)-\tau_{C}(y)\leq 0\,; \ee hence \be
\sup_{y\in\bR^2}\Big\{\,(y|x)-\tau_{C}(y)\,\Big\}=0\,, \ee that
is $x\in A(n)$.

Conversely, suppose that $x\in A(n)\cap C$. Since $x\in A(n)$ we
have \be (x|n)=\tau_{C}(n)\,. \ee Since $x\in C$ we have for all
$z$ \be \tau_{C}(z+n)\geq (z+n|x)\,. \ee Therefore \be
\tau_{C}(z+n)\geq (n|x) + (z|n) =\tau_{C}(n) +(z|n)\,. \ee \qed
As a consequence of
Theorem \ref{geometry2} the support plane $A(n)$ is regular if
and only if $\tau_{C}$ is differentiable at $n$, $\|n\|=1$. On
the other hand the regular points of $\partial C$ are
characterized by the points of $\partial C$ where the normal is
well--defined (there is a unique tangent plane at those points).
We can partition the boundary  $\partial C$ of $C$ into three
sets.

\begin{enumerate} \item $x$ is regular and the support plane
$A(n)$, $n$ the unit normal vector to $C$ at $x$, is regular.
The support function $\tau_{C}$ is differentiable at $n$. \item
$x$ is regular and the support plane $A(n)$, $n$ the unit normal
vector to $C$ at $x$, is not regular. The support function
$\tau_{C}$ is not differentiable at $n$. The {\bf facet} of
(unit) normal $n$ is given by the subdifferential of the support
function $\tau_{C}$ at $n$, \be A(n)\cap C=\partial \tau_{C}\,.
\ee \item $x$ is not regular. Such a point corresponds to a {\bf
corner} of $C$, where the normal is not well--defined. There are
at least two different support planes.
\end{enumerate}

{\bf Remarks:} 1. The dual function of $\tau_{C}$ is \be
\tau_{C}^*(x):=\sup_{y\in\bR^2}\Big\{\,(x|y)-\tau_{C}(y)\,\Big\}\,.
\ee In our case the dual function $\tau_{C}^*$ is the {\bf
indicator function} of $C$: \be \tau_{C}^*(x)=\cases{0 & if
$x\in C$,\cr \infty & if $x\not\in C$.\cr} \ee (See proof of
Theorem \ref{geometry2}.) The dual function of $\tau_{C}^*$ is
$\tau_C$, \be
\tau_{C}^{**}(x):=\sup_{y\in\bR^2}\Big\{\,(x|y)-\tau_{C}^*(y)\,\Big\}
=\tau_{C}(x)\,. \ee

2. There is another interesting function associated to $C$, the
{\bf polar} of $C$, $\tau_{C}^{\circ}$, \be
\tau_{C}^{\circ}(x):=
\sup_{n\in\bR^2\,\|n\|=1}{(x|n)\over\tau_{C}(n)}\,. \ee The
convex body $C$ can be expressed as \be
C=\Big\{x\in\bR^2:\,\tau_{C}^{\circ}(x)\leq 1\Big\}\,, \ee and
its boundary as \be \partial
C=\Big\{x\in\bR^2:\,\tau_{C}^{\circ}(x)=1\Big\}\,. \ee A point
of $\partial C$ is regular if and only if $\tau_{C}^{\circ}$ is
differentiable at $x$.

\subsection{ Isoperimetric inequality}

Let $C$ be a compact convex body in $\bR^2$ and $\tau_{C}$ its
support function. Let $\partial V$ be a rectifiable curve in
$\bR^2$, which is the boundary of an open set $V$. The curve
$\partial V$ is oriented in the usual way;  $n(x)$ denotes the
exterior unit normal vector at a point $x\in\partial V$. We
define a functional 
\be 
\cF_C(\partial V):=\int_{\partial
V}\,\tau_C(n(s))ds\,. 
\ee 
If $\gamma :[0,t]\ra \bR^2$ is a
parametrization of the boundary $\partial V$, then, using the
homogeneity property of $\tau_C$, we can write the
functional as 
\be
\cF_{C}(\partial V):=
\int_{0}^{t}\,\tau_{C}\Big((\dot{\gamma}(2)(u),-\dot{\gamma}(1)(u))
\Big)du\,, 
\ee
where $\dot{\gamma}(s)$ is the derivative of
$\gamma$ with respect to $s$. The functional is always positive.
Indeed, by a suitable translation $a$ we can suppose that the
translated set $C'=C+a$ has zero as interior point, and
consequently its support function $\tau_{C'}$ is strictly
positive at $x\not= 0$. Since \be
\tau_{C'}(x)=\tau_{C}(x)+(x|a)\,, \ee and \be \int_{\partial
V}\,(n(s)|a)ds=0\,, \ee the value of the functional does not
depend on $a$.

\begin{thm}[Generalized isoperimetric inequality]
\label{thmisoineq}

Let $C$ be a compact convex body in $\bR^2$. Let $V$ be an open
set in $\bR^2$ such that its boundary  is a
rectifiable\footnote{ The theorem is valid under less
restrictive assumptions; see e.g. \cite{F}} curve $\partial V$.
The Lebesgue measure of $C$ and $V$ is denoted by $|C|$ and
$|V|$. Then \be\label{isoineq} \cF_{C}(\partial V)\geq 2
|C|^{1/2}|V|^{1/2}\,. \ee Equality holds in (\ref{isoineq}) if
and only if $V$ equals, up to dilation and translation, the set
$C$. \end{thm}

{\bf Remarks:} 1. The set $V$ may have several connected
components.

2. The main ingredient of Dinghas' proof \cite{D} is to express
the functional $\tau_{C}$ in a geometrical manner as \be
\cF_{C}(\partial V)=\lim_{\varepsilon\downarrow 0}
{|V+\varepsilon C|-|V|\over \varepsilon}\,, \ee where
$V+\varepsilon C$ denotes the set \be \bigcup_{x\in
V}\,(x+\varepsilon C)\,, \ee and $\varepsilon C=\{\varepsilon
x:\,x\in C\}$. The isoperimetric inequality follows by applying
Brunn-Minkowski inequality to $V+\varepsilon C$, \be
|V+\varepsilon C|^{1/2}\geq |V|^{1/2}+|\varepsilon C|^{1/2}\,.
\ee

3. When $C$ is the unit ball we recover the classical
isoperimetric inequality.

4. The minimum of the functional for open sets $V$ with
$|V|=|C|$ can easily be computed, \be
\min_{V:\,|V|=|C|}\cF_{C}(\partial V)=2\cdot|C|\,. \ee \bigskip

In addition to the isoperimetric inequality, the stability of
the minimum can be controlled in a rather strong sense by the
(generalized) {\bf Bonnesen's inequalities}. Let \be r(\partial
V)=\sup\Big\{r:\,r\cdot C+x\subset V\; \hbox{for some
$x\in\bR^2$}\Big\} \ee and \be R(\partial
V)=\inf\Big\{R:\,R\cdot C+x\supset V\; \hbox{for some
$x\in\bR^2$}\Big\}\,. \ee Then \bea {\cF_{C}(\partial
V)-\Bigl[\cF_{C}(\partial V)^2 -4|C|\cdot|V|\Bigr]^{1/2}\over
2|C|} &\leq& r(\partial V)\leq R(\partial V)\\ &\leq&
{\cF_{C}(\partial V)+\Bigl[\cF_{C}(\partial V)^2
-4|C|\cdot|V|\Bigr]^{1/2}\over 2|C|}\,. \nonumber \eea

\subsection{Variational problem}\label{problem}

Let $r_1, r_2\in\bN$ be given and define the rectangle $Q$, \be
Q:=\{x=(x(1),x(2))\in\bR^2:\,-r_1\leq x(1)\leq r_1\,;\, 0\leq
x(2)\leq 2r_2\}\,. \ee Let $\beta>\beta_c$ and
$\widehat{\tau}:\bR^2\ra\bR$ be the surface tension. Let
$S^1=\{n\in\bR^2:\,\|n\|=1\}$; we define on $Q\times S^1$ a
function $\t$, \be\label{tt} \t(x;n):=\cases{\tau^--\tau^+ &if
$x(2)=0$ and $n=n_w$,\cr \widehat{\tau}(n)&otherwise.\cr} \ee
We define a functional as above, by
setting \be\label{wtt} \W_{\t}(\partial V):=\int_{\partial
V}\,\t(s;n(s))\,ds\,. \ee Notice that $\tau^--\tau^+ $ is
negative when $h$ is negative. However, as long as
$\tau^--\tau^+ >-\widehat{\tau}(n_w)$, then the functional is
positive. This is an easy consequence of the elementary
monotonicity principle stated in subsection \ref{solution}.

Let us give the interpretation of $\W_{\t}(\gamma)$ in our
setting. We consider a macroscopic droplet of the $-$ phase
immersed in the $+$ phase. The boundary of the droplet is
$\partial V$.  Suppose first that $x(2)\not=0$ and $\|n\|=1$.
Then $\t(x;n)$ is interpreted as the surface tension of an
interface perpendicular to $n$, passing through $x$. The case
$x(2)=0$ and $n=n_w$ corresponds to the situation where the
macroscopic droplet is in contact with the wall. In that case
\be \tau^{-}-\tau^{+} \ee is the change in the free energy due
to the presence of the droplet on the wall (the bulk phase is
the $+$ phase).

{\bf Variational problem VP:} {\sl Suppose that the bulk phase
in $Q$ is the $+$ phase and that the $-$ phase occupies a set
$V$ such that $|V|$ is a fraction $\alpha$ of the volume of $Q$,
$|V|=\alpha |Q|$. $V$ is not necessarily connected, but we assume
that it is open and that its boundary $\partial V$ is a
rectifiable curve. Find the optimal set $C$, $|C|=\alpha|Q|$,
such that $C$ minimizes the functional $\W_{\t}(\partial V)$,
$|V|=\alpha|Q|$.}

{\bf Remarks:} 1.  In \cite{PV} we consider the same variational
problem. Its formulation is slightly different, because we have
introduced as fundamental quantities the dual quantities, the
decay--rates of the two--point functions.

2. There is a simple way of showing that the solution $C$ is a
convex body. Let us consider the case where
$\t(x,n)=\widehat{\tau}(n)$. Since $ \widehat{\tau}$ is convex
and positively homogeneous,
we have for any parametrized curve $\gamma:\,[a,b]\ra\bR^2$, by
Jensen's inequality \be {1\over b-a}
\int_a^b\widehat{\tau}\Big((\dot{\gamma}(2)(s),-\dot{\gamma}(1)(s))
\Big)ds \geq \ee $$
\widehat{\tau}\Bigl(\Bigl({\gamma(2)(b)-\gamma(2)(a)\over b-a},
-{\gamma(1)(b)-\gamma(1)(a)\over b-a}\Bigr)\Bigr)\,. $$
Therefore one decreases the value of the functional each time we
replace some part of the curve between two points by the
straight segment between these two points. Since the box $Q$ is
convex, for every open set $V\subset Q$ its convex enveloppe
${\rm conv}\,V\subset Q$, and $|{\rm conv}\,V|\geq|V|$. On the
other hand \be \W_{\t}(\partial V)\geq \W_{\t}\Big(\partial({\rm
conv}\,V)\Big)\,. \ee

\subsubsection{Solution of the variational problem}
\label{solution}

We give the explicit solution in some special cases. We refer to
\cite{KP} for other cases.  We consider cases where only the
effect of the bottom horizontal wall is important. This amounts
to consider cases where the amount of $-$ phase is not too
large. (Theorem \ref{mainthm} is valid without such
restrictions.)

The function $\widehat{\tau}$ can be interpreted as the support
function of a compact convex set, which we denote by
$C_{\widehat{\tau}}$, \be
C_{\widehat{\tau}}:=\Big\{x\in\bR^2:\,(x|y)\leq
\widehat{\tau}(y)\;\forall y\Big\}\,. \ee Indeed, the dual
function $\widehat{\tau}^{*}$ is the indicator function of the
set $C_{\widehat{\tau}}$, so that the support function of
$C_{\widehat{\tau}}$, which the dual of the indicator function
is $\widehat{\tau}^{**}=\widehat{\tau}$, because
$\widehat{\tau}$  is convex. The compact convex body
$C_{\widehat{\tau}}$ is called {\bf Wulff shape} in the physics
literature.

{\bf Case 1:} $\tau^--\tau^+=\widehat{\tau}(n_w)$.

In that case $\t$ is independent of the variable $x$, so that
\be \t(x,n)=\widehat{\tau}(n)\,. \ee Let us fix the volume of
the set $V$ occupied by the $-$ phase to be $\alpha |Q|$,
$0<\alpha<1$. Ignoring for a moment the constraint that
$V\subset Q$, the solution of the variational problem is given
by Theorem \ref{thmisoineq}. It is a Wulff shape of volume
$\alpha |Q|$, $\lambda \cdot C_{\widehat{\tau}}$, with \be
\lambda={\alpha |Q|\over|C_{\widehat{\tau}}|}\,. \ee Whenever a
translate of $\lambda\cdot C_{\widehat{\tau}}$ can be put inside
$Q$, this is a solution of the variational problem. When there
is no translate of $\lambda\cdot C_{\widehat{\tau}}$ which can
be put inside $Q$, then the constraint that $V\subset Q$
modifies the shape of the optimal set $V$ (see \cite{KP}).

{\bf Case 2:} $|\tau^--\tau^+|<\widehat{\tau}(n_w)$.

\begin{figure}[t]
\centerline{\psfig{figure=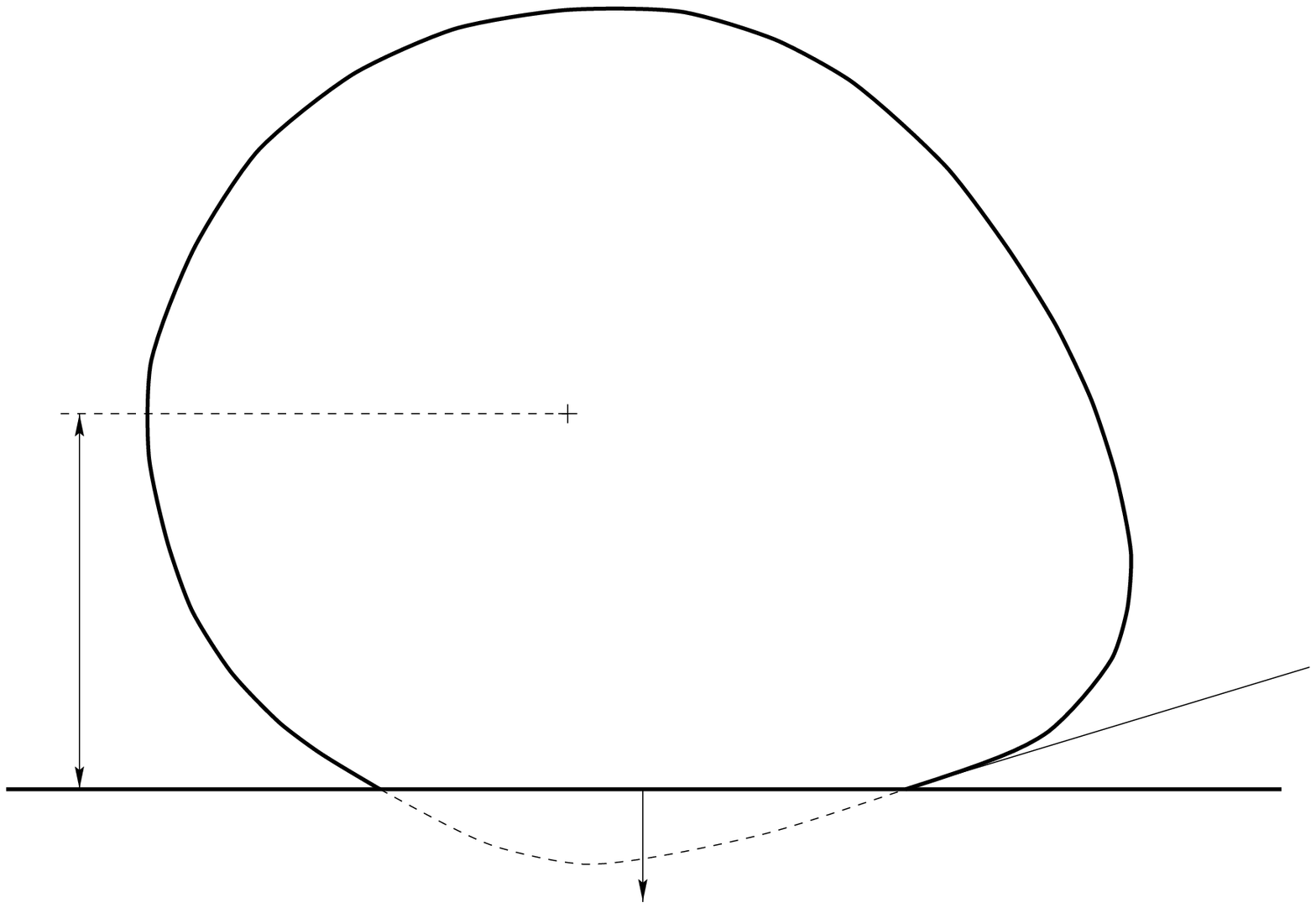,height=50mm}}
\figtext{ 
\writefig	7.75	3.32	{$O$}
\writefig	7.50	0.87	{$n_w$}
\writefig	3.45	2.10	{$\tau^--\tau^+$} } 
\legende{Winterbottom
construction.} \label{fig_winterbottom} \end{figure}

Let us first ignore the constraint that $V\subset Q$. Notice
that the problem without the constraint $V\subset Q$ is
scale--invariant. Then the solution is given by a Winterbottom
shape of volume $\alpha |Q|$. The {\bf Winterbottom shape} is by
definition the convex body (see Fig. \ref{fig_winterbottom})
\be
C_{\t}:=\widehat{C}\cap\{x:\,(x|n_w)\leq\tau^--\tau^+\}\,. 
\ee

To
prove the optimality of the Winterbottom shape we use Theorem
\ref{thmisoineq} and the following {\bf monotonicity principle}
\cite{KP}. Suppose that we can find a convex body $C$, such that
the following two conditions (\ref{A}) and (\ref{B}) are
verified. If we replace in the definition of the functional
$\W_{\t}$ the function $\t$ by the support function $\tau_{C}$,
then \be\label{A} \W_{\t}(\partial V)\geq \W_{\tau_{C}}(\partial
V)\,, \ee and for $V=C$, \be\label{B} \W_{\t}(\partial C)=
\W_{\tau_{C}}(\partial C)\,. \ee Then we have \be\label{1}
\W_{\t}(\partial V)\geq 2  |C|^{1/2}|V|^{1/2} \ee and
\bea\label{2} {\W_{\t}(\partial V)-\Bigl[\W_{\t}(\partial V)^2
-4|C|\cdot|V|\Bigr]^{1/2}\over 2|C|} &\leq& r(\partial V)\leq
R(\partial V)\\ &\leq& {\W_{\t}(\partial
V)+\Bigl[\W_{\t}(\partial V)^2 -4|C|\cdot|V|\Bigr]^{1/2}\over
2|C|}\,. \nonumber \eea The proof of (\ref{1}) is an immediate
consequence of Theorem \ref{thmisoineq} applied to the convex
set $C$. The proof of (\ref{2}) is an immediate consequence of
the monotonicity of the real function \be s\mapsto
s-(s^2-D)^{1/2} \ee for $|s|\geq \sqrt{D}$. In our case
$C=C_{\t}$.  It is immediate that condition (\ref{B}) is
verified. To show (\ref{A}) we  notice that $\t^*$ is the
indicator function of $C_{\t}$. Therefore its support function
satisfies \be \t(n)\geq\t^{**}(n)=\tau_{C}(n)\,. \ee

\begin{figure}[htb]
\centerline{\psfig{figure=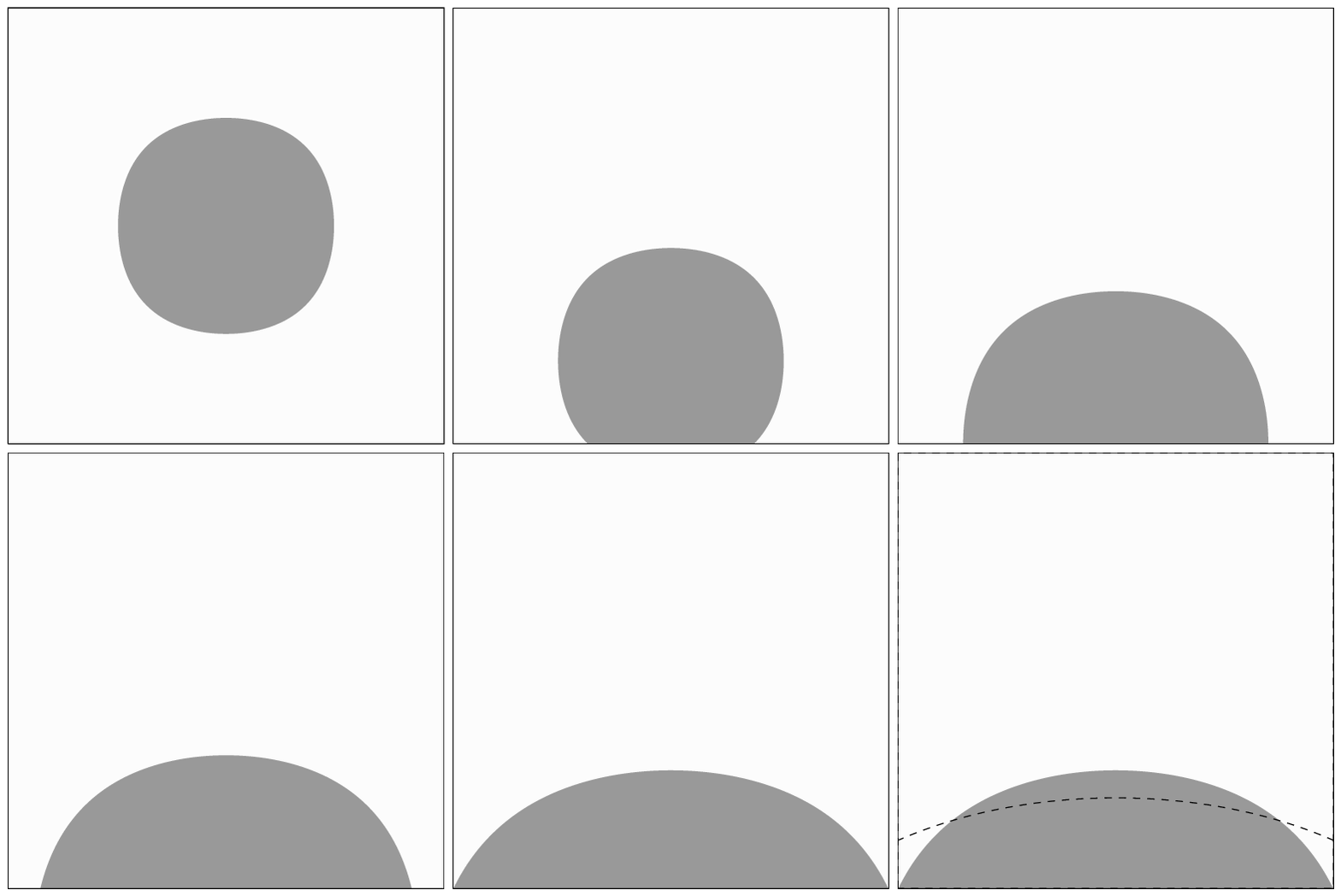,height=80mm}}
\figtext{ 
\writefig	2.3	8.0	{a}
\writefig	6.3	8.0	{b} 
\writefig	10.3	8.0	{c}
\writefig	2.3	4.0	{d} 
\writefig	6.3	4.0	{e}
\writefig	10.3	4.0	{f} } 
\legende{a: $h>h^*$; b: $0<h<h^*$; c:
$h=0$; d, e, f: a sequence of droplets for decreasing values of
$-h^*<h<0$. The droplet spreads until it begins to touch the
vertical sides of the box (e). Further reduction of the magnetic
field does not modify the shape of the droplet, but makes it
unstable in the sense that the removal of the vertical walls
would result in a spreading of the droplet. f: the dashed line
shows only a part of the droplet which would be obtained by
removing the walls when $0>h>-h^*$.} \label{fig_droplet}
\end{figure}

\begin{figure}[htb]
\centerline{\psfig{figure=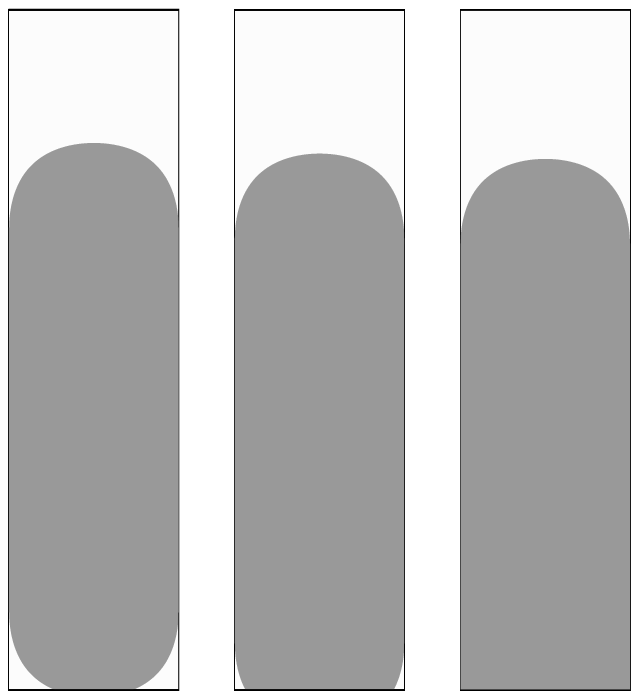,height=70mm}}
\figtext{ 
\writefig	5.2	6.93	{a} 
\writefig	7.50	6.93	{b}
\writefig	9.77	6.93	{c} 
} 
\legende{Sequence of big droplets in
a tube for decreasing value of the magnetic field. a,b: The
upper part of the droplet has the Wulff shape, while the lower
part has the winterbottom shape; they are joined by a rectangle
such that the total volume is conserved; c: The droplet
completely wets the lower wall, the droplet is build up from a
half Wulff shape and a rectangle. This situation holds as long
as $-h^*<h \leq 0$.} \label{fig_bigdroplet} \end{figure}

Any translate of the Winterbottom shape, which is contained in
$Q$,  is a solution of the variational problem. When there is no
translate  which is inside $Q$, then the constraint that
$V\subset Q$ effectively modifies the shape of the optimal set
$V$. We want to discuss one simple situation of that kind (see Fig.
\ref{fig_droplet} and \ref{fig_bigdroplet} for illustrations).
Suppose that $\alpha$ is chosen so that a translate of the
Winterbottom shape of volume $\alpha|Q|$ exists inside $Q$
whenever $h\geq 0$. Let $h(\alpha)$ be the smallest value of $h$
such that a translate of the Winterbottom shape of volume
$\alpha|Q|$ exists inside $Q$. Notice that necessarily
$h(\alpha)>- h^*$, and that the value of $h(\alpha)$ is a
function of the box $Q$. Then for any $h\leq h(\alpha)$ the
solution of the problem is the same as the solution of the
problem for $h=h(\alpha)$: the box prevents the macroscopic
droplet to spread out. For $h\leq h(\alpha)$  the optimal shape
of the droplet of volume $\alpha|Q|$ is not the corresponding
Winterbottom shape. 

\begin{figure}[htb]
\centerline{\psfig{figure=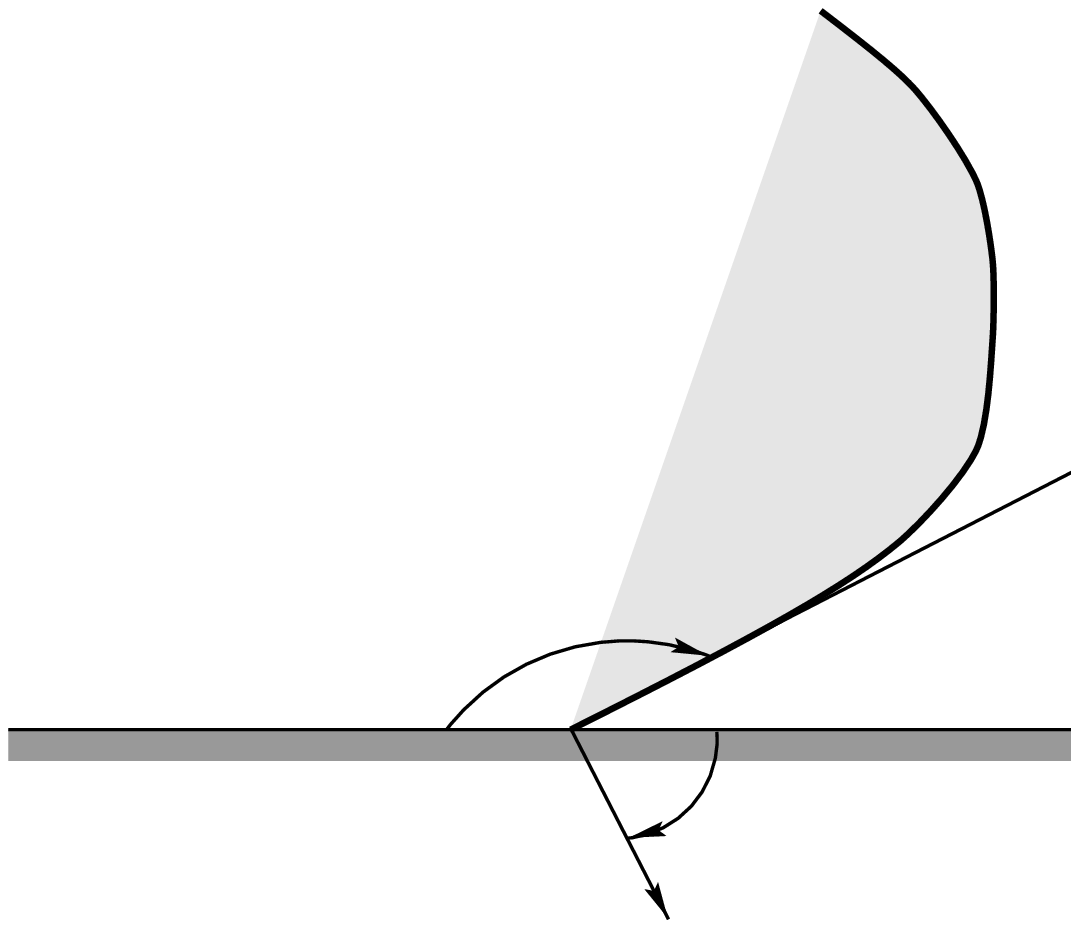,height=50mm}}
\figtext{ 
\writefig	8.85	0.9	{$-\varphi$}
\writefig	8.05	0.9	{$n$} 
\writefig	7.60	2.15	{$\theta$} 
}
\legende{Angles $\theta$ and $\varphi$} \label{fig_angles}
\end{figure} 
\bigskip {\bf Remark:} The Young--Herring relation, giving the
contact angle $\theta$ of the macroscopic droplet with the wall
in terms of the surface tension and surface free energies, can
be easily derived from Theorem \ref{geometry2}. Let us
parametrize the unit vectors in $\bR^2$ by an angle $\varphi$ as
in figure \ref{fig_angles}. Let $n$ be the normal to the interface
defining the contact angle $\theta$. Theorem \ref{geometry2}
asserts that \be {\rm grad}\widehat{\tau}(n)(2)=-(\tau^-
-\tau^+)\,. \ee This is exactly Young--Herring relation. Indeed,
in polar coordinates the surface tension is \be
\widehat{\tau}(r,\varphi):=r\cdot\widehat{\tau}(n(\varphi))\,,
\ee so that, using $x(2)=r\sin\varphi$ and
$\varphi+\theta=\pi/2$, we get
($\widehat{\tau}(1,\varphi)\equiv\widehat{\tau}(\varphi)$) \bea
{\rm grad}\widehat{\tau}(n)(2)&=&
\sin\varphi\,\widehat{\tau}(\varphi)+ \cos\varphi\,{d\over
d\varphi}\widehat{\tau}(\varphi)\\ &=&
\cos\theta\,\widehat{\tau}(\theta)- \sin\theta\,{d\over
d\theta}\widehat{\tau}(\theta) \nonumber\\ &=& -(\tau^-
-\tau^+)\,.\nonumber \eea 


\section{Canonical ensemble and macroscopic
droplet}\label{canens.} \setcounter{equation}{0}

To study the macroscopic droplet of fixed volume we introduce a
canonical ensemble. There is some freedom in doing this, in the
sense that we can fix the total magnetization up to
fluctuations, which are negligible when measured at the scale of
the volume. In subsection \ref{canonical} we define the
canonical states and in subsection \ref{droplet} we consider the
limit of the lattice spacing going to zero. We show that in this
limit there is a droplet of the $-$ phase immersed in the $+$
phase and that the shape of the droplet is given by the
variational problem of subsection \ref{problem}. This completes
the mathematical theory of wetting in the 2D Ising model in
terms of the Gibbs states. We always consider the case of $+$
boundary condition.

\subsection{Canonical states}\label{canonical}

We define the canonical states at finite volume. Let
$\beta>\beta_c$ and $-m^*<m<m^*$. Let $c=1/4-\delta>0$, with
$\delta>0 $. We introduce the event \be A(m;c):=\{\omega:\,
|\,\sum_{t\in\Lambda_L}\omega(t)-m|\Lambda_L|\,| \leq
|\Lambda_L|\cdot L^{-c}\}\,. \ee

The {\bf canonical state} in $\Lambda$, with $+$ boundary
condition and parameter $m$, is the conditional state
\be\label{cann}
\bk{\,\cdot\,|\,m}_L^+(\beta,h):=\bk{\,\cdot\,|\,A(m;c)}_L^+(\beta,h)\,.
\ee To understand the canonical state (\ref{cann}) we must
control the large deviations of the magnetization in the state
$\bk{\,\cdot\,}^+_L$. Although for typical set of configurations
with respect to  $\bk{\,\cdot\,}_L^+$  the length of the contours
is small \footnote{For any $\beta>\beta_c$ there exits
$K(\beta)$ so that $ \lim_{L\ra\infty}\bk{\,\{\exists\;\hbox{a
contour of length} \;\geq K\ln L\}\,}_L^+=0\,. $ (Lemma 5.6 in
\cite{PV}.) }, because we impose here a specific magnetization
$m\not =m^*$ there is always at least one large contour in each
typical set of configurations of the canonical state. It is
therefore natural to distinguish between large and small
contours. Let $B(0;[L^{\delta}])$ be the square box 
\be
B(0;[L^{\delta}]):=\{t\in\bR^2:\,|t(j)|\leq [L^{\delta}] \;,\;
j=1,2\}\,. 
\ee
We say that a contour $\gamma$ is {\bf small} if
there exits a translate of the box $B(0;[L^{\delta}])$ which
contains $\gamma$. Otherwise the contour is ${\bf large}$. We
sum over small contours and the large contours are treated by a
coarse--graining method. Theorems 11.1 and 11.2 in \cite{PV} give
a detailled description of a set of typical configurations in terms of 
large contours. One result of this analysis is the exact computation
of the large deviations of the magnetization for the Gibbs state
$\bk{\,\cdot\,}_L^+(\beta,h)$, together with a 
control of the speed of convergence.

\begin{thm}\label{mainthm} 
Assume that $\beta>\beta_c$,
$h\in\bR$, $-m^*(\beta)<m<m^*(\beta)$ and $c:=1/4-\delta$,
$\delta>0$. 
Let $\W^{*}_{\t}(m)$ be defined by 
\be
\W^{*}_{\t}(m):=\inf\Big\{\W_{\t}(\cC):\,\cC\subset Q\,,\, {\rm
vol\,}\cC={m^*-m\over 2m^*}|Q|\Big\}\,. 
\ee
Then for any $\eta <\delta$ and $L$ large enough
\be 
|\,{1\over L}\ln P_L^{{\rm h}}[A(m;c)]+\W^{*}_{\t}(m)\,|\leq
O(L^{\eta-\delta}) \,; 
\ee
the probability is computed with the measure 
$\bk{\,\cdot\,}_L^+(\beta,h)$.
\end{thm}

\subsection{Macroscopic droplet}\label{droplet}

In this last subsection we consider the limit of the lattice
spacing going to zero. We suppose that $\beta>\beta_c$,
$h\in\bR$ and we choose the $+$ boundary condition. The
probability measure in this section is always the canonical Gibbs state
$\bk{\,\cdot\,|m}_L^{+}(\beta,h)$ defined in subsection
\ref{canonical} with $-m^*<m<m^*$ and $c=1/4-\delta>0$ fixed.
We do the analysis in the box
$\Lambda_L(r_1,r_2)$ and at the end we scale everything by $1/L$,
i.e. we take the limit of the lattice spacing going to zero.

Let $C\subset\bZ^2$; {\bf the empirical magnetization in} $C$ is
\be 
m_{C}(\omega):={1\over |C|}\sum_{t\in C}\sigma(t)(\omega)\,.
\ee 
Let $0<a<1$; we introduce a grid $\cL(a)$
in $\Lambda_L$ made of cells which are
translates of the square box 
\be
B(0;[L^{a}])=\{t\in\bR^2:\,|t(j)|\leq [L^{a}] \;,\;
j=1,2\}\,. 
\ee
The value 
of $a$ is  close to $1$. In most of
the cells the empirical magnetization is close to $m^*$ or
$-m^*$ with high probability. For each cell of the grid $\cL(a)$ 
we compute the empirical magnetization $m_C(\omega)$. Then we scale all lengths by
$1/L$, so that after scaling the box $\Lambda_L$ 
is the rectangle $Q$. For each $\omega$ we define a magnetization profile
$\rho_{L}(x;\omega)$ on $Q$,
\be
\rho_{L}(x;\omega):=m_{C}(\omega)\; \hbox{if $Lx\in C$}\, 
\ee
where $Lx$ is the point $x\in Q$ scaled by $L$ and $C$ a cell
of the grid $\cL(a)$.\\
The set of macroscopic droplets at equilibrium is 
\be
\cD(m):=\{\,\cV\subset Q:\,|\cV|={m^*-m\over 2m^*}|Q|\,,\,
\W_{\t}(\partial\cV)=\W_{\t}^*(m)\,\}\,. 
\ee 
For each
$\cV\in\cD(m)$ we have a magnetization profile, 
\be
\rho_{\cV}(x):=\cases{m^*& if $x\in Q\backslash \cV$,\cr -m^*&
$x\in \cV$.\cr} 
\ee
 Let $f$ be a real--valued function on $Q$;
we set 
\be
d_1(f,\cD(m)):=\inf_{\cV\in\cD(m)}\int_{Q}dx\,|\,f(x)-\rho_{\cV}(x)\,|\,.
\ee 
The main theorem (Theorem 12.2 in \cite{PV}) is

\begin{thm}\label{distance} Let $\beta>\beta_c$, $h\in\bR$,
$-m^*<m<m^*$, $c=1/4-\delta>0$. 
Let $\bk{\,\cdot\,|\,m\,}_L^{+}(\beta,h)$  be the canonical Gibbs
state with $+$ boundary condition. Then there exists a positive
function $\overline{\varepsilon}(L)$ such that
$\lim_{L\ra\infty}\overline{\varepsilon}(L)=0$ and for $L$ large
enough \be {\rm
Prob}[\,\{\,d_1(\rho_{L}(\,\cdot\,;\omega),\cD(m))\leq
\overline{\varepsilon}(L)\,\}\,]\geq
1-\exp\{-O(L^{\kappa})\}\,. 
\ee 
\end{thm}

This theorem gives a complete description of the wetting phenomenon
in the canonical state, making the connection between a microscopic
approach with a conditional state and the macroscopic variational
problem giving the shape of a macroscopic droplet at equilibrium.
The two approaches with grand canonical and canonical ensembles
give of course the same information about the occurence of the wetting
transition. However, for this surface phenomenon they are not equivalent,
but complementary.


\newpage

\begin{thebibliography}{99}

\bibitem{A} Abraham D.B.: {\sl Solvable model with a
roughening transition for a planar Ising ferromagnet} Phys. Rev.
Lett. {\bf 44}, 1165--1168 (1980).


\bibitem{ABCP} Alberti G., Bellettini G., Cassandro M.,
Presutti E.: {\sl Surface tension in Ising systems with Kac
potentials} to appear in J. Stat. Phys. (1996).




\bibitem{BLP1} Bricmont J., Lebowitz J.L., Pfister C.-E.:
{\sl On the surface tension of lattice systems.} Annals of the
New York Academy of Sciences {\bf 337}, 214--223 (1980).

\bibitem{C} Cahn J.W.: {\sl Critical point wetting} J. Chem.
Phys. {\bf 66}, 3667-3672 (1977).






\bibitem{D} Dinghas A.: {\sl Uber einen geometrischen Satz
von Wulff fur die Gleichtgewichtsform von Kristallen.}
Zeitschrift fur Kristallographie {\bf 105}, 301--314 (1944).

\bibitem{DP} Dacorogna B., Pfister C.-E.: {\sl Wulff theorem
and best constant in Sobolev inequality.} J. Math. Pures  Appl.
{\bf 71} 97--118 (1992).



\bibitem{DKS} Dobrushin R.L., Koteck\'y R., Shlosman S.:
{\sl Wulff construction: a global shape from local interaction.}
AMS translations series (1992).

\bibitem{E} Eggleston  {\sl Convexity} Cambridge University
Press, Cambridge (1977).

\bibitem{F} Fonseca I.: {\sl The Wulff theorem revisited}
Proc.R.Soc. Lond. A {\bf 432}, 125--145 (1991).

\bibitem{FP1} Fr\"ohlich J., Pfister C.-E.: {\sl
Semi--infinite Ising model I. Thermodynamic functions and phase
diagram in absence of magnetic field.} Commun. Math. Phys. {\bf
109}, 493--523 (1987).

\bibitem{FP2} Fr\"ohlich J., Pfister C.-E.: {\sl
Semi--infinite Ising model II. The wetting and layering
transitions.} Commun. Math. Phys. {\bf 112}, 51--74 (1987).

\bibitem{FP3} Fr\"ohlich J., Pfister C.-E.: {\sl The
wetting and layering transitions in the half--infinite Ising
model.}  Europhys. Lett. {\bf 3}, 845--852 (1987).

\bibitem{I1} Ioffe D.: {\sl Large deviations for the 2D
Ising model: a lower bound without cluster expansions.} J. Stat.
Phys. {\bf 74} 411--432 (1994).

\bibitem{I2} Ioffe D.: {\sl Exact large deviation bounds up
to $T_c$ for the Ising model in two dimensions.} Prob. Th. Rel.
Fields. {\bf 102} 313--330 (1995).

\bibitem{KP} Koteck\'y R., Pfister C.-E.: {\sl Equilibrium
shapes of crystals attached to walls.} J. Stat. Phys. {\bf 76},
419-445 (1994).

\bibitem{LP} Lebowitz J.L., Pfister C.-E.: {\sl Surface
tension and phase coexistence.} Phys. Rev. Lett. {\bf 46},
1031--1033 (1981).

\bibitem{MS1} Minlos R.A., Sinai Ya.G.: {\sl The phenomenon
of phase separation at low temperatures in some lattice models
of a gas I.} Math. USSR-- Sbornik {\bf 2}, 335--395 (1967).

\bibitem{MS2} Minlos R.A., Sinai Ya.G.: {\sl The phenomenon
of phase separation at low temperatures in some lattice models
of a gas II.} Trans. Moscow Math. Soc. {\bf 19}, 121--196
(1968).

\bibitem{MW} McCoy B.M., Wu T.T.: {\sl The Two--dimensional
Ising Model.} Harvard University Press, Cambridge, Massachusetts
(1973).

\bibitem{O1} Osserman R.: {\sl The isoperimetric inequality}
Amer. Math. Soc. {\bf 84}, 1182--1238 (1978).

\bibitem{O2} Osserman R.: {\sl Bonnesen-style isoperimetric
inequalities} Amer. Math. Monthly {\bf 86}, 1--29 (1979).

\bibitem{P} Patrick A.: private communication (1996).

\bibitem{Pf1} Pfister C.-E.: {\sl Interface and surface
tension in Ising model. In Scaling and self--similarity in
physics.} ed. J. Fr\"ohlich, Birkh\"auser, Basel, pp. 139--161
(1983).

\bibitem{Pf2} Pfister C.-E.: {\sl Large deviations and
phase separation in the two--dimensional Ising model.} Helv.
Phys. Acta {\bf 64}, 953--1054 (1991).

\bibitem{PP} Pfister C.-E., Penrose O.: {\sl Analyticity
properties of the surface free energy of the Ising model}
Commun. Math. Phys. {\bf 115}, 691--699 (1988).

\bibitem{PV} Pfister C.-E., Velenik Y.: {\sl Large
deviations and boundary effects for the 2D Ising model.}
Preprint (1996). To appear in Prob. Th. Rel. Fields.

\bibitem{RW} Rotman C., Wortis M.: {\sl Statistical
mechanics of equilibrium crystal shapes: Interfacial phase
diagrams and phase transitions} Phys. Rep. {\bf 103}, 59--79
(1984).

\bibitem{T} Taylor J.E.: {\sl Some crystalline variational
techniques and results} Ast\'erisque {\bf 154-155}, 307--320
(1987).

\bibitem{Wi} Winterbottom W.L.: {\sl Equilibrium shape of a
small particle in contact with a foreign substrate} Acta
Metallurgica {\bf 15}, 303--310 (1967).

\bibitem{Wu} Wulff G.:  {\sl Zur Frage der Geschwindigkeit
des Wachstums and der Aufl\"osung der Kristallfl\"achen.}
Zeitschrift fur Kristallographie  {\bf 34}, 449-530  (1901).

\bibitem{Z} Zia R.K.P.: {\sl Anisotropic surface tension and
equilibrium crystal shapes} in Progress in Statistical Physics,
C.K. Hu, ed., World Scinetific, Singapore (1988), 303--357.

\end{thebibliography}
\end{document}